\documentclass{aa}
\pdfoutput=1

\usepackage[version=3]{mhchem}
\usepackage[varg]{txfonts}
\usepackage{natbib,siunitx,booktabs,multirow,pgfplots,hyperref}
\bibpunct{(}{)}{;}{a}{}{,}
\pgfplotsset{compat=newest}
\usepgfplotslibrary{external}
\DeclareSIUnit\Kms{\milli\kelvin\km\per\s}
\DeclareSIUnit\kms{\km\per\s}
\DeclareSIUnit\kgs{\kg\per\s}
\DeclareSIUnit\mols{{molec.}\,s^{-1}}
\defcitealias{2010A&A...518L.149B}{Paper~I}
\defcitealias{2010A&A...518L.150H}{Hartogh et al. 2010}
\defcitealias{2007P&SS...55.1058B}{Biver et al. 2007}

\newcommand{\wbsho}{11.4(35)}
\newcommand{\hrsho}{16.2(46)}
\newcommand{\wbsv}{-71(40)}
\newcommand{\wbshv}{-240(90)}
\newcommand{\wbsvv}{-10(70)}
\newcommand{\wbsdv}{-230(120)}
\newcommand{\hrsv}{-77(67)}
\newcommand{\hrshv}{-230(120)}
\newcommand{\hrsvv}{-150(66)}
\newcommand{\hrsdv}{-80(140)}
\newcommand{\wbsqho}{1.7(5)e27}
\newcommand{\hrsqho}{2.3(7)e27}
\newcommand{\qho}{2.0(5)e27}

\newcommand{\wbsn}{<13}
\newcommand{\hrsn}{<17}
\newcommand{\wbsqn}{<1.5e27}
\newcommand{\hrsqn}{<1.9e27}
\newcommand{\qnh}{<1.5e27}

\newcommand{\qpacs}{\SI{<1.4e28}{\mols}}
\newcommand{\qspire}{\SI{<4e28}{\mols}}
\newcommand{\ratran}{\texttt{ratran}}
\newcommand{\herschel}{{\it Herschel}}
\newcommand{\akari}{{\it AKARI}}
\newcommand{\spitzer}{{\it Spitzer}}
\newcommand{\christensen}{C/2006~W3 (Christensen)}
\newcommand{\chris}{C/2006~W3}
\newcommand{\trans}{$J_{K_\mathrm{a} K_\mathrm{c}}\ (1_{10}\text{--}1_{01})$}
\newcommand{\nh}{$J_K\ (1_{0}\text{--}0_{0})$}
\newcommand{\xne}{x_{n_\mathrm{e}}}
\newcommand{\rh}{r_\mathrm{h}}

\pgfplotsset{
  spectrum style/.style={
    axis x line*=bottom,
    const plot mark mid,
    no markers,
    extra x ticks       = 0,
    extra x tick labels = ,
    extra y ticks       = 0,
    extra y tick labels = ,
    extra tick style  = { grid = major },
    minor x tick num=4,
    minor y tick num=4,
    xmin=-10, xmax=10,
    scaled y ticks={real:1e-3},
    ytick scale label code/.code={},
    xlabel={$v$ [\si{\kms}]},
    ylabel={$T_\mathrm{mB}$ [\si{\milli\K}]},
  },
  lsb style/.style={
    xlabel={$\nu_\mathrm{LSB}$ [GHz]},
    ymin=0, ymax=1,
    axis x line*=top,
    axis y line=none,
    x dir=reverse,
    minor x tick num=4,
  },
  map style/.style={
    width=8cm,
    enlargelimits=false,
    axis equal image,
    axis on top,
    x dir=reverse,
    xlabel={$\Delta\alpha_{2000}$ [\arcsec]},
    ylabel={$\Delta\delta_{2000}$ [\arcsec]}
  },
  profile style/.style={
    only marks,
    minor x tick num=4,
    xlabel={$\rho$ [arcsec]},
    ylabel={$S_\mathrm{B}$ [\si{Jy.arcsec^{-2}}]},
    xmin=1, xmax=60,
    clip marker paths,
  },
}

\begin{document}

\title{\herschel{} observations of gas and dust in comet
\christensen{} at  5 AU from the Sun\thanks{\herschel{} is an ESA space
observatory with science instruments provided by European-led Principal
Investigator consortia and with important participation from NASA.}
}

\author{M.~de~Val-Borro\inst{\ref{mps}}\fnmsep\inst{\ref{pu}}
  \and D.~Bockel\'ee-Morvan\inst{\ref{meudon}}
  \and E.~Jehin\inst{\ref{liege}}
  \and P.~Hartogh\inst{\ref{mps}}
  \and C.~Opitom\inst{\ref{liege}}
  \and S.~Szutowicz\inst{\ref{src}}
  \and N.~Biver\inst{\ref{meudon}}
  \and J.~Crovisier\inst{\ref{meudon}}
  \and D.~C.~Lis\inst{\ref{caltech}}
  \and L.~Rezac\inst{\ref{mps}}
  \and Th.~de~Graauw\inst{\ref{alma}}
  \and D.~Hutsem\'ekers\inst{\ref{liege}}
  \and C.~Jarchow\inst{\ref{mps}}
  \and M.~R.~Kidger\inst{\ref{herschel}}
  \and M.~K\"uppers\inst{\ref{rosetta}}
  \and L.~M.~Lara\inst{\ref{iaa}}
  \and J.~Manfroid\inst{\ref{liege}}
  \and M.~Rengel\inst{\ref{mps}}
  \and B.~M.~Swinyard\inst{\ref{ral}}\fnmsep\inst{\ref{ucl}}
  \and D.~Teyssier\inst{\ref{herschel}}
  \and B.~Vandenbussche\inst{\ref{leuven}}
  \and C.~Waelkens\inst{\ref{leuven}}
  }

\titlerunning{Gas and dust activity in \christensen{}}

\institute{
  Max-Planck-Institut f\"ur Sonnensystemforschung,
    Justus-von-Liebig-Weg~3, 37077 G\"{o}ttingen, Germany\label{mps}\\
    \email{hartogh@mps.mpg.de}
  \and Department of Astrophysical Sciences, Princeton University,
    Princeton, NJ 08544, USA\label{pu}\\
    \email{valborro@princeton.edu}
  \and LESIA, Observatoire de Paris, CNRS, UPMC, Universit\'e
    Paris-Diderot, 5 place Jules Janssen, 92195 Meudon,
    France\label{meudon}
  \and Institut d'Astrophysique et de G\'eophysique, Universit\'e de
    Li\`ege, 4000 Li\'ege, Belgium\label{liege}
  \and Space Research Centre, Polish Academy of Sciences, Bartycka 18A,
    00-716 Warsaw, Poland\label{src}
  \and California Institute of Technology, Pasadena, CA 91125,
    USA\label{caltech}
  \and Joint ALMA Observatory, Alonso de C\'ordova 3107, Vitacura,
    Santiago, Chile\label{alma}
  \and \herschel{} Science Centre, ESAC, European Space Agency, 28691
    Villanueva de la Ca\~nada, Madrid, Spain\label{herschel}
  \and Rosetta Science Operations Centre, ESAC, European Space Agency,
    28691 Villanueva de la Ca\~nada, Madrid, Spain\label{rosetta}
  \and Instituto de Astrof\'isica de Andaluc\'ia (CSIC), Glorieta de la
    Astronomía s/n, 18008 Granada, Spain\label{iaa}
  \and RAL Space, Rutherford Appleton Laboratory, Chilton, Didcot OX11
    0QX, UK\label{ral}
  \and Department of Physics and Astronomy, University College London,
    Gower St., London WC1E 6BT, UK\label{ucl}
  \and Instituut voor Sterrenkunde, Katholieke Universiteit Leuven,
    Belgium\label{leuven}
  }

\date{Received 15 January 2014 / Accepted 26 February 2014}

\abstract
  {Cometary activity at large heliocentric distances is thought to be
  driven by outgassing of molecular species more volatile than water
  that are present in the nucleus.  The long-period comet \christensen{}
  was an exceptional target for a detailed study of its distant gaseous
  and dust activity.
  }
  {We aimed to measure the \ce{H2O} and dust production rates in
  \christensen{} with the \herschel{} Space Observatory at a
  heliocentric distance of $\sim$ 5 AU and compared these data with
  previous post-perihelion \herschel{} and ground-based observations at
  $\sim$ 3.3 AU from the Sun.}
  {We have searched for emission in the \ce{H2O} and \ce{NH3}
  ground-state rotational transitions, \trans{} at \SI{557}{\giga\hertz}
  and \nh{} at \SI{572}{\giga\hertz}, simultaneously, toward comet
  \christensen{} with the Heterodyne Instrument for the Far Infrared
  (HIFI) onboard \herschel{} on UT 1.5 September 2010.  Photometric
  observations of the dust coma in the \SIlist{70;160}{\um} channels
  were acquired with the Photodetector Array Camera and Spectrometer
  (PACS) instrument on UT 26.5 August 2010.
  }
  {A tentative 4-$\sigma$ \ce{H2O} line emission feature was
  found in the spectra obtained with the HIFI wide-band and
  high-resolution spectrometers, from which we derive a water production
  rate of \SI{\qho}{\mols}.  A 3-$\sigma$ upper limit for the ammonia
  production rate of \SI{\qnh}{\mols} is obtained taking into account
  the contribution from all hyperfine components.  The dust thermal
  emission was detected in the \SIlist{70;160}{\um} filters, with a more
  extended emission in the blue channel.  We fit the radial dependence
  of the surface brightness with radially symmetric profiles for the
  blue and red bands. The dust production rates, obtained for a dust
  size distribution index that explains the fluxes at the photocenters
  of the \SIlist{70;160}{\micro\m} PACS images, lie in the range
  \SIrange{70}{110}{\kg\per\s}.  Scaling the CO production rate measured
  post-perihelion at \SIlist{3.20;3.32}{AU}, these values correspond to
  a dust-to-gas production rate ratio in the range 0.3--0.4.
  }
  {The blueshift of the water line detected by HIFI suggests preferential
  emission from the subsolar point.  However, it is also possible that
  water sublimation occurs in small ice-bearing grains that are emitted
  from an active region on the nucleus surface at a speed of \SI{\sim
  0.2}{\kms}.  The dust production rates derived in August 2010 are
  roughly one order of magnitude lower than in September 2009,
  suggesting that the dust-to-gas production rate ratio remained
  approximately constant during the period when the activity became
  increasingly dominated by CO outgassing.
  }


\maketitle

\section{Introduction}

Comets are small solar system bodies with a wide range of orbital
periods. They spend most of their time in the cold outer regions of the
solar system, beyond the distance at which water condenses into ice
grains---the so-called snow line.  Therefore, they contain pristine
material that reflects the chemical composition in the early solar
nebula.  Comets are composed of a loose aggregation of volatile icy
materials and refractory particles.  Observations of gas production in
comets provide a unique opportunity to constrain the composition of
cometary ices and to characterize a classification scheme based on their
relative chemical abundances
\citep{1995Icar..118..223A,2002EM&P...90..323B,2004come.book..391B,2009EM&P..105..267C}.
Studying the processes responsible for ice sublimation in comets is also
important in the understanding of the activity and thermal properties of
cometary nuclei.  In addition, the simultaneous imaging of the thermal
emission from dust particles released from the nucleus in two
wavelengths bands can constrain important properties of the dust coma,
such as the dust size distribution index and dust production rate
\citep[e.g.,][]{1990ApJ...365..738J,2010A&A...518L.149B}.  Therefore,
observations of cometary dust and gas can provide clues about the
physical conditions in the solar nebula, and help to establish a link
between the materials in the parent interstellar cloud and cometary
nuclei \citep[see e.g.,][]{2004come.book..115E}.

Cometary activity at small heliocentric distances is driven mainly by
sublimation of water molecules from the nucleus with \ce{CO2}/\ce{H2O}
and \ce{CO}/\ce{H2O} mixing ratios that show a great diversity and range
from $\sim$ 0.01--0.1 in most cometary atmospheres
\citep[e.g.,][]{2011IAUS..280..261B,2013A&A...559A..48D}.  Although
\ce{H2O} is generally the primary component of the nucleus, the
sublimation of water is inefficient at $\rh$ larger than
\SIrange{3}{4}{AU} and the activity is dominated by molecules such as
\ce{CO2} and \ce{CO} that are more volatile than water \citep[see
e.g.,][]{2010A&A...518L.149B,2012Icar..220..277M}. Therefore, the
evolution of the production rate of these molecular species with
heliocentric distance provides important clues for understanding the
structure and composition of cometary nuclei
\citep[e.g.,][]{2002EM&P...90....5B}.

The \herschel{} Space Observatory \citep{2010A&A...518L...1P} has proven
to be the most sensitive facility for directly observing water emission
in distant comets and studying the chemical composition of cometary
material \citep[see
e.g.,][]{2011Natur.478..218H,2012A&A...544L..15B,2013ApJ...774L...3L}.
Water emission has been detected in the distant comet
29P/Schwassmann-Wachmann~1 \citep{2010DPS....42.0304B}. Moreover, recent
attempts have been made to directly detect sublimating water using
\herschel{} in comets \christensen{} \citep{2010A&A...518L.149B} and
C/2012 S1 (ISON) \citep{2013A&A...560A.101O} at large heliocentric
distances, and also in active objects orbiting within the main asteroid
belt, that is, main-belt comets \citep[MBCs; see
e.g.,][]{2012A&A...546L...4D,2013ApJ...774L..13O}.

Comet \object{C/2006 W3 (Christensen)} is a long-period comet that was
discovered in November 2006 at a distance of 8.6 AU from the Sun. It
passed perihelion on 6 July 2009 at a heliocentric distance of 3.13 AU.
As noted previously, water ice sublimation is expected to be ineffective
at large heliocentric distances and, owing to the low outgassing rate
that requires very sensitive observations, it has only been directly
detected previously in comet \christensen{} at infrared wavelengths in
the pre-perihelion observations at \SIlist{3.66;3.13}{AU} by the
\akari{} satellite on UT 21.1 December 2008 and UT 16.8 June 2009,
respectively \citep{2012ApJ...752...15O}.

At the time of the previous \herschel{} post-perihelion observations of
comet \christensen{} at 3.3 AU from the Sun \citep[][hereafter
\citetalias{2010A&A...518L.149B}]{2010A&A...518L.149B} the Heterodyne
Instrument for the Far Infrared \citep[HIFI;][]{2010A&A...518L...6D} was
not available due to a single event upset in the memory of the Local
Oscillator Control Unit microcontroller caused by a cosmic-particle
impact.  Had HIFI observations been obtained at that time, water
emission in the coma would have been readily detected.  Nonetheless, an
upper limit to the water production rate of \qpacs{} was derived from
spectroscopic observations with the Photodetector Array Camera and
Spectrometer \citep[PACS;][]{2010A&A...518L...2P}  on UT 8.8 November
2009, and an upper limit of \qspire{} was obtained from observations
with the Spectral and Photometric Imaging REceiver
\citep[SPIRE;][]{2010A&A...518L...3G} on UT 6.6 November 2009.
Additionally, the dust size distribution and dust production rate were
calculated from the spectral energy distribution measured by PACS
\citepalias[see][]{2010A&A...518L.149B}.  Complementary
production rate measurements of several species (namely CO, \ce{CH3OH},
HCN, \ce{H2S}, and OH) were obtained from the ground in pre-perihelion
observations with the Nan\c{c}ay radio telescope and post-perihelion
observations with the Institut de Radioastronomie Millim\'etrique (IRAM)
30-m telescope \citepalias[][]{2010A&A...518L.149B}.

Here we present the analysis of more \herschel{} outbound observations
of comet \christensen{} at 5 AU from the Sun.  The paper is structured
as follows: in Sect.~\ref{sec:observations}, we summarize the
observations performed with HIFI and PACS, and the data reduction
method.  Section~\ref{sec:results} presents the results of the data
analysis.  In Sect.~\ref{sec:resultshifi}, we compare the gas production
rate obtained by HIFI with measurements by other facilities at various
heliocentric distances.  The derived dust production rates using a dust
thermal emission model are compared with the values obtained in the
previous PACS observations in Sect.~\ref{sec:resultspacs}.  Finally, we
discuss the main results of this work in Sect.~\ref{sec:discuss} and
present the conclusions in Sect.~\ref{sec:conclusions}.

\begin{table*}
  \caption{HIFI and PACS observing circumstances of comet
    \christensen{}.}
  \label{tbl:log}
  \centering
  \begin{tabular}{c S[table-format = 3] c
		  S[table-format = 10,
		  table-space-text-post = \textsuperscript{\emph{h}}]
		  S[table-format = 2.1]
		  S[table-format = 3]
		  c c
		  c c
		  S[output-decimal-marker = {\fdg}]
		  }
    \toprule
    Date\tablefootmark{\emph{a}} &
    {OD\tablefootmark{\emph{b}}} &
    Inst. & {ObsID} & {Exp.} &
    {Angle\tablefootmark{\emph{c}}} &
    {Scan size} &
    {Speed\tablefootmark{\emph{d}}} &
    $\rh$\tablefootmark{\emph{e}} &
    $\Delta$\tablefootmark{\emph{f}} &
    {$\phi$\tablefootmark{\emph{g}}}\\
    (yyyy-mm-dd.ddd) & & & & {(\si{\minute})} &
    {(\si{\degree})} & {($\si{\arcmin} \times \si{\arcmin}$)} &
    {(\si{\arcsec\per\s})} &
    (AU) & (AU) & {(\si{\degree})}\\
    \midrule
    2009-11-01.836 & 171 & PACS & 1342186621\textsuperscript{\emph{h}} & 9.4 & 135 & $9\farcm9\times7\farcm4$ & 10 &3.33 & 3.56 & 16.31\\
2009-11-01.843 & 171 & PACS & 1342186622\textsuperscript{\emph{h}} & 9.4 & 45 & $9\farcm9\times7\farcm4$ & 10 &3.33 & 3.56 & 16.31\\
2010-08-26.532 & 469 & PACS & 1342203478 & 20.6 & 45 & $3\farcm0\times1\farcm7$ & 20 &4.96 & 4.54 & 11.22\\
2010-08-26.547 & 469 & PACS & 1342203479 & 20.6 & 135 & $3\farcm0\times1\farcm7$ & 20 &4.96 & 4.54 & 11.22\\
2010-09-01.529 & 475 & HIFI & 1342204014 & 48.0 &  &  &  &5.00 & 4.68 & 11.45\\

    \bottomrule
  \end{tabular}
  \tablefoot{
  \tablefoottext{\emph{a}}{UT mid-date of the observation with
  fractional days.}
  \tablefoottext{\emph{b}}{\herschel{} operational day.}
  \tablefoottext{\emph{c}}{Orientation angle of the scan map with
  respect to the detector array.}
  \tablefoottext{\emph{d}}{Slewing speed of the satellite along the scan
  line legs.}
  \tablefoottext{\emph{e}}{Heliocentric distance.}
  \tablefoottext{\emph{f}}{Distance to \herschel.}
  \tablefoottext{\emph{g}}{Solar phase angle (Sun--\chris--Earth).}
  \tablefoottext{\emph{h}}{Data from \citetalias{2010A&A...518L.149B}.}
  }
\end{table*}

\section{Observations}
\label{sec:observations}

The \herschel{} Space Observatory is a 3.5-m telescope built and
operated by the European Space Agency (ESA) covering the far-infrared
and submillimetre wavelength ranges \citep{2010A&A...518L...1P}.  Comet
\christensen{} (hereafter referred to as \chris{}) was observed with
PACS and HIFI, two of the focal-plane instruments onboard \herschel{},
on UT 26.5 August and 1.5 September 2010, respectively, when the object
was at $r_\mathrm{h} \sim 5.0$ AU and a distance of $\Delta$ = 4.5–4.7
AU from the spacecraft.  The observations were performed within the
framework of the \herschel{} guaranteed-time key program ``Water and
related chemistry in the solar system'' \citep{2009P&SS...57.1596H}.
These data complement previous outbound observations of comet \chris{}
obtained four months post-perihelion with PACS during the science
demonstration phase on UT 1.8 and 8.8 November 2009 and with SPIRE on UT
6.6 November 2009, that have been reported previously in
\citetalias{2010A&A...518L.149B}.

Table~\ref{tbl:log} summarizes the observing circumstances and distances
to comet \chris{} during the observations obtained with the HIFI and
PACS instruments.  We used the ephemeris provided by JPL's HORIZONS
online solar system data
service\footnote{\url{http://ssd.jpl.nasa.gov/?horizons}} to obtain the
position and solar position angle of the comet with respect to the
satellite \citep{1996DPS....28.2504G}.

The pointing information of \herschel{} observations acquired in the
period between operational days (ODs) 320 and 761 suffered from a shift
in the reconstructed astrometry of up to \SI{8}{\arcsec} due to an
introduced change in the temperature of the star-tracker.  This effect
was modified by a new model that was uploaded to the satellite on OD
762.  For the PACS and HIFI observation of \chris{} obtained on ODs 469
and 475, respectively, new pointing products have been generated by
reprocessing the data with the new astrometry generated with the
\herschel{} Interactive Processing Environment (HIPE) v10.3.0 that
corrects the pointing offset problem.

\subsection{HIFI observations}

Double-sideband (DSB) heterodyne systems such as HIFI are sensitive to
two frequency ranges located on either side of the local oscillator
frequency.  Using HIFI, we aimed to simultaneously detect the
ground-state rotational transitions \trans{} of ortho-\ce{H2O} and \nh{}
of ortho-\ce{NH3} at \SIlist{557;572}{\giga\hertz}, in the lower and
upper sidebands (LSB and USB) of band 1b, with a total integration time
of \SI{48}{\minute}.  The observation was performed in the single-point
frequency-switching observing mode without a reference position on the
sky that maximizes the on-target integration time, although it
introduces strong standing waves in the baseline. These standing waves
have to be removed and are a source of additional systematic
uncertainties.  A frequency throw of 94.5 MHz was applied in this
observation.  We simultaneously used the wide-band spectrometer (WBS)
and the high-resolution spectrometer (HRS) with frequency resolutions of
\SI{1.1}{\mega\hertz} and \SI{120}{\kilo\hertz}, after resampling  to a
uniform frequency grid.  Nonetheless, note that both the WBS and HRS
spectra are oversampled and the spectrometers have a spacing of the
frequency grid of $\sim$ \SIlist{500;60}{\kilo\hertz}, respectively.  In
the HIFI observation the \ce{H2O} and \ce{NH3} emission line frequencies
were targeted in `subband' 4 of the WBS, whereas the HRS has a single
subband in the high-resolution observing mode.

The data reduction was carried out with the standard HIFI pipeline
v10.3.0 using the HIPE
software package to obtain calibrated level-2 products
\citep{2010ASPC..434..139O}.  We corrected the temperature scale
using the HIFI forward efficiency, i.e., the fraction of radiation
received from the forward hemisphere of the beam as compared to the
total radiation received by the antenna, of 0.96.  Additionally a main
beam efficiency of 0.75 in band 1b has been taken into account to
correct for the fraction of power collected in the main Gaussian beam of
the telescope with respect to the total power
\citep[see e.g.,][]{2012A&A...537A..17R}.

\subsection{PACS observations}

In photometer mode, the PACS instrument simultaneously images two of its
three filters centered on \SIlist{70;100;160}{\um} (referred to as red,
green, and blue bands) that cover the 60--85 \si{\um}, 85--125 \si{\um},
and 125--210 \si{\um} ranges.  Two bolometer arrays provide a field of
view of $1\farcm75 \times 3\farcm5$ in each of the bands.

The PACS maps presented here were taken in the red and blue bands with
two orthogonal scanning directions with respect to the detector array,
using the medium slewing speed of the spacecraft along parallel lines of
\SI{20}{\arcsecond/\s} \citepalias[as compared with the slow-scan mode
  of \SI{10}{\arcsecond/\s} for the observation obtained during the
\herschel{} science verification phase presented
in][]{2010A&A...518L.149B}.  The integration time was \SI{20.6}{\minute}
for each scanning direction. In the earlier PACS observations from
November 2009, we used three scan legs with a 9\farcm9 length and a
$\sim$ 2\farcm5 leg separation, while the more recent observations have
five scan legs with a 3\arcmin{} length and 20\arcsec{} separation.  The
pixel sizes are $6\farcs4 \times 6\farcs4$ and $3\farcs2 \times
3\farcs2$ for the red and blue channels, but for maps smaller than
5\arcmin{} in size, as in the case of the \chris{} observations from
August 2010, the PACS photometry pipeline resamples the images to pixel
sizes of 1\arcsec{} and 2\arcsec{} in the blue and red bands.  The PACS
data from November 2009 and August 2010 were reduced with the HIPE
pipeline version v10.3.0.

\section{Data analysis}
\label{sec:results}

\subsection{HIFI data analysis}
\label{sec:resultshifi}

The data reduction for the HIFI data follows the workflow described for
example in \citet{2012A&A...539A..68B} and \citet{2012A&A...546L...4D}.
After processing the raw data to calibrated level-2 products with the
HIFI pipeline, the frequency-switched data are folded by averaging the
original spectrum with an inverted copy that is shifted by the local
oscillator throw of \SI{94.5}{\MHz}
\citep[e.g.,][]{1997A&AS..124..183L}.  A strong baseline ripple is
present, introduced by multiple standing waves in the
frequency-switching observing mode.  To obtain a reliable estimate of
the noise in the emission-free part of the spectrum, the baseline has to
be removed, which was accomplished by fitting a linear combination of
sine waves using the Lomb-Scargle periodogram technique
\citep{1976Ap&SS..39..447L,2010ApJS..191..247T}.  The spectrum can also
be processed using an empirical mode-decomposition (EMD) method
\citep[see][and Rezac et al.\ in preparation for a quantitative
comparison of Lomb-Scargle and EMD techniques with other
procedures]{2014A&A...563A...4R}.  We used a masking window of (-1,1)
\si{\kms} around the transition frequency in each of the orthogonal
polarizations spectra to fit the standing waves in the baseline.

The procedure used to remove the residual standing waves of
frequency-switched observations introduces an additional systematic
error source of about 50\% of the root mean square (rms) scatter
of the baseline-subtracted spectrum \citep[see
e.g.,][]{2012A&A...544L..15B,2013ApJ...774L..13O}. Moreover, an
additional uncertainty, similar to that arising from the baseline
removal, is related to the flux calibration error associated with the
beam efficiency error, calculated from Mars mapping observations,
and sideband gain ratio between the upper and lower sidebands
\citep[see][]{2012A&A...537A..17R}.  We have adopted a conservative
value of 15\% flux error in band 1b resulting from a combination of
these uncertainties \citep[see e.g.,][]{HIFIcal}.  All these systematic
errors were considered in the derived line intensity and production rate
uncertainties we present below.

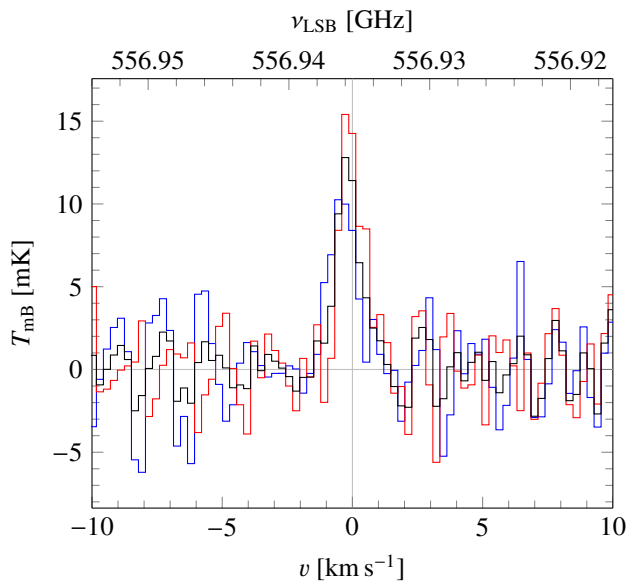
\begin{figure}
  \centering
  \begin{tikzpicture}
    \begin{axis}[lsb style,
      xmin=556.917, xmax=556.954,
      xtick={556.91,556.92,556.93,556.94,556.95},
      ]
    \end{axis}
    \begin{axis}[spectrum style]
      \addplot[blue] table [x index=1,y index=2]
	{1342204014_WBS-H_fluxcal.dat};
      \addplot[red] table [x index=1,y index=2]
	{1342204014_WBS-V_fluxcal.dat};
      \addplot[black] table [x index=1,y index=2]
	{1342204014_WBS_fluxcal.dat};
    \end{axis}
  \end{tikzpicture}
  \caption{
    Folded spectra for the horizontal polarization mixer (blue line),
    vertical polarization mixer (red line), and averaged spectrum (black
    line) of the ortho-water \trans{} line at 556.936 GHz acquired with
    the WBS on UT 1.5 September 2010.  The fitted baselines are obtained
    by masking a (-1, 1) \si{\kms} window around the rest frequency of
    the water line and using a linear combination of sine functions.
    The vertical axis is the calibrated main-beam brightness
    temperature, the upper horizontal axis is the lower sideband
    frequency, and the lower horizontal axis shows the Doppler velocity
    with respect to the nucleus rest frame.
  }
  \label{fig:wbs}
\end{figure}

Figure~\ref{fig:wbs} shows the folded and baseline-subtracted WBS
spectra of the \trans{} transition of \ce{H2O} measured in horizontal
and vertical polarizations (hereafter H and V).  We averaged the H and V
polarization data together to improve the signal-to-noise
ratio (S/N) of the emission feature and used the same method to remove
the standing waves in the baseline (black line shown in
Fig.~\ref{fig:wbs}).  A tentative detection of the \ce{H2O} line is
present in the folded spectra at the position of the transition rest
frequency for both polarizations.  However, there are similar features
in the spectrum introduced by the standing waves in the baseline ripple.
The line feature is present in the folded H and V polarization spectra
with a similar intensity and S/N of 3.5, although not exactly at the
same frequency; the H spectrum is blueshifted by about
\SI{\wbshv}{\m\per\s}.

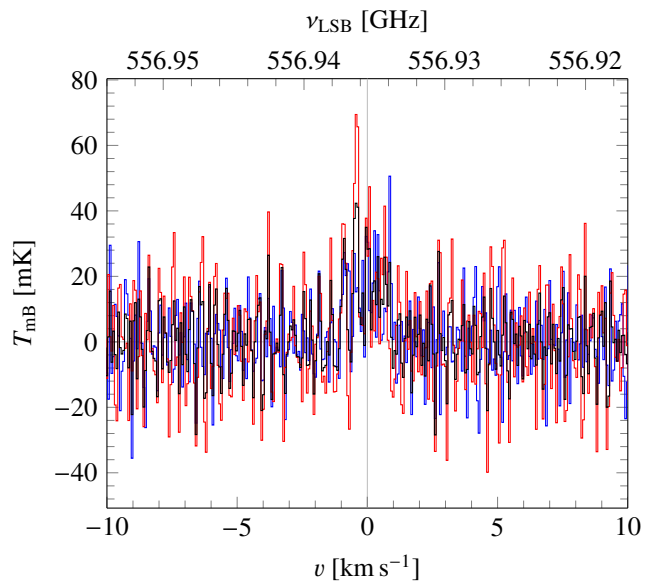
\begin{figure}
  \centering
  \begin{tikzpicture}
    \begin{axis}[lsb style,
      xmin=556.917, xmax=556.954,
      xtick={556.91,556.92,556.93,556.94,556.95},
      ]
    \end{axis}
    \begin{axis}[spectrum style]
      \addplot[blue] table [x index=1,y index=2]
	{1342204014_HRS-H_fluxcal_H2O.dat};
      \addplot[red] table [x index=1,y index=2]
	{1342204014_HRS-V_fluxcal_H2O.dat};
      \addplot[black] table [x index=1,y index=2]
	{1342204014_HRS_fluxcal_H2O.dat};
    \end{axis}
  \end{tikzpicture}
  \caption{
    Folded HRS spectra for the horizontal polarization mixer (blue line),
    vertical polarization mixer (red line), and averaged spectrum (black
    line) of the ortho-water \trans{} line at 556.936 GHz
    obtained on UT 1.5 September 2010.  The fitted baselines are obtained
    using the same method as for the WBS data.  The vertical axis is the
    calibrated main-beam brightness temperature, the upper horizontal
    axis is the lower sideband frequency, and the lower horizontal axis
    shows the Doppler velocity with respect to the nucleus rest frame.
  }
  \label{fig:hrs}
\end{figure}

Using the same method as in the WBS spectrum to fit and subtract the
baseline, an emission feature appears in the HRS data that is consistent
with the line area detected by the WBS within 1-$\sigma$ uncertainties
including the systematic component from the baseline subtraction.
Figure~\ref{fig:hrs} shows the reduced spectra of the water \trans{}
line measured by the HRS.  Note that to use the spectrometer's native
frequency resolution and to increase the S/N of the spectra, we
resampled over two adjacent channels with a rectangular window function,
which gives an effective spectral resolution of \SI{120}{\kilo\hertz}.
After removing the baseline by a linear combination of sinusoids, the
integrated intensities of the line are \SIlist{\wbsho;\hrsho}{\Kms} in
the main-beam brightness temperature scale for WBS and HRS.  The
agreement in the line intensities within uncertainties is to be expected
since it is the same signal coming from the mixer including the standing
waves, that is multiplexed to the WBS and HRS. This also confirms that
the spectral feature has not been introduced by either of the two
backends.

The line centroids in the HRS data with the original frequency
resolution are shifted by \SIlist{\hrshv;\hrsvv}{\m\per\s} with respect
to the rest frequency of the \ce{H2O} transition in the H and V spectra,
respectively, compared to a blueshift of
\SIlist{\wbshv;\wbsvv}{\m\per\s} for the WBS.  The difference between
the H and V line centroids in the HRS spectra, \SI{\hrsdv}{\m\per\s}, is
smaller than in the estimation in the WBS data, \SI{\wbsdv}{\m\per\s},
while they agree within $\sim$ 1-$\sigma$ uncertainties.  Although the
HRS data has higher frequency resolution than the WBS data, the S/N in
the emission line in the HRS spectra is very low to allow a good
estimate of the difference between the H and V line velocity offsets.

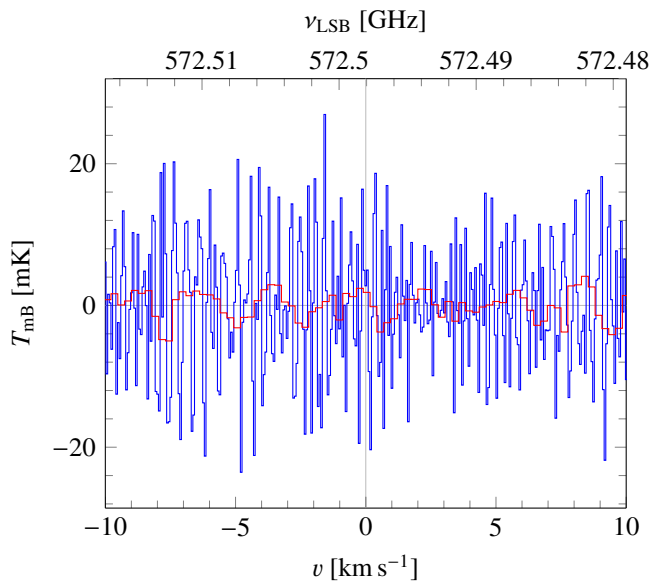
\begin{figure}
  \centering
  \begin{tikzpicture}
    \begin{axis}[lsb style,
      xmin=572.479, xmax=572.517,
      ]
    \end{axis}
    \begin{axis}[spectrum style]
      \addplot table [x index=1,y index=2] {1342204014_HRS_NH3.dat};
      \addplot table [x index=1,y index=2] {1342204014_WBS_NH3.dat};
    \end{axis}
  \end{tikzpicture}
  \caption{
    Folded averaged spectra of the \nh{} transition of \ce{NH3} at
    \SI{572.498}{\giga\hertz} obtained with the HRS (blue line) and WBS
    (red line) on UT 1.5 September 2010.  The vertical axis is the
    calibrated main-beam brightness temperature, the upper horizontal
    axis shows the upper sideband frequency, and the lower horizontal
    axis shows the Doppler velocity with respect to the nucleus rest
    frame.
    }
  \label{fig:nh3}
\end{figure}

We simultaneously searched for \ce{H2O} and \ce{NH3} emission in the
HIFI observation of \chris{} in the LSB and USB.  The averaged spectra
around the ammonia transition observed with the two HIFI backends and
centered on the barycentric position of the hyperfine component
frequencies are shown in Fig.~\ref{fig:nh3}.  We removed the
residual standing waves in the calibrated spectra using the same
procedure as described for the \ce{H2O} emission line. There is no
indication of an \ce{NH3} line emission in either the WBS or HRS data.
The derived 3-$\sigma$ upper limits to the \ce{NH3} line intensity
including systematic uncertainties are \SIlist{\wbsn;\hrsn}{\Kms} for
the WBS and HRS spectra.

\subsubsection{Origin of the difference in the H- and V-polarization
spectra}

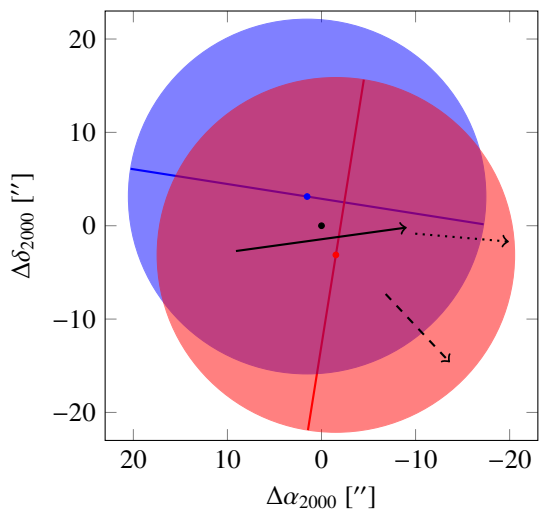
\begin{figure}
  \centering
  \begin{tikzpicture}
    \begin{axis}[
      xmin=-23, xmax=23, ymin=-23, ymax=23,
      axis equal image,
      x dir=reverse,
      xlabel={$\Delta\alpha_{2000}$ [\arcsec]},
      ylabel={$\Delta\delta_{2000}$ [\arcsec]}
      ]
      \draw[blue,thick] (axis cs:-17.237, 0.154)--(axis cs:20.295, 6.0989);
      \draw[red,thick] (axis cs:1.44, -21.893)--(axis cs:-4.5, 15.639);
      \draw[blue,opacity=0.5,fill=blue,fill opacity=0.5]
      (axis cs:1.5286,3.1266) circle [radius=19];
      \draw[red,opacity=0.5,fill=red,fill opacity=0.5]
      (axis cs:-1.5286,-3.1266) circle [radius=19];
      \draw[blue,fill=blue] (axis cs:1.5286,3.1266) circle [radius=0.3];
      \draw[red,fill=red] (axis cs:-1.5286,-3.1266) circle [radius=0.3];
      \draw[black,fill=black] (axis cs:0, 0) circle [radius=0.3];
      \draw[->,thick] (axis cs:9.09195,-2.71135)--(axis cs:-9.0739,-0.18898);
      \draw[->,dotted,thick] (axis cs:-9.964,-0.851) --
	(axis cs:-19.927,-1.702);
      \draw[->,dashed,thick] (axis cs:-6.820,-7.314) --
	(axis cs:-13.640,-14.627);
    \end{axis}
  \end{tikzpicture}
  \caption{Weighted average of the point spread functions (PSFs) for the
    horizontal polarization mixer (blue circle) and vertical
    polarization mixer (red circle) obtained at various times during the
    HIFI observation towards comet \christensen. The orientation of the
    orthogonal polarizations is shown by the thick blue and red lines
    (H and V).  The trajectory of the comet nucleus projected on the sky
    during the HIFI integration of \SI{48}{\minute} is shown by the
    solid line according to the HORIZONS ephemeris.  The projected
    direction toward the Sun and projected velocity vector of the comet
    with respect to the origin are indicated by the dotted and dashed
    arrows.  The figure is centered on the average synthetic beam
    aperture located in the center of the two polarization beams and
    shown by the black point.
  }
  \label{fig:pointing}
\end{figure}

The difference in the \ce{H2O} line position for the orthogonal
polarizations in the HIFI WBS data can be partly explained by the
pointing offset between H and V polarization spectra of $\sim$ 6\farcs6
in band 1b.  A comparison of the pointing offset of the H and V beams
with the full width at half maximum (FWHM) of the averaged beams is
shown in Fig.~\ref{fig:pointing} ($\sim$ \SI{38}{\arcsec} at these
frequencies, corresponding to an aperture of $\sim$ \SI{130000}{\km} at
a distance of 4.68 AU from \herschel{}).  For \herschel{} observations
of solar system objects, the tracking mode is activated to follow their
motion on the sky. In the case of comet \chris{} the HORIZONS ephemeris
was used to determine the tracking coefficients for the instrument
boresight that are used in the pointing commands.  For HIFI's band 1,
the orientation of the polarization grids is such that the H
polarization is basically parallel to the ecliptic plane and the V
polarization lies perpendicular to it, in the case of sources close to
the ecliptic plane.  The projected trajectory of the comet nucleus on
the sky during the observation passes farther away from the center of
the averaged H beams (mean angular distance $\sim$ 4\farcs8) than from
the center of the averaged V beams (mean angular distance $\sim$
2\farcs3) as shown in Fig.~\ref{fig:pointing}.

As noted previously, a pointing correction due to a warm star-tracker
during the HIFI observation was taken into account by the HIPE pipeline
that resulted in an offset of
$\cos(\delta_\mathrm{e}) \times
(\delta_\mathrm{e}-\delta_\mathrm{o}) = 1\farcs0$ and
$(\alpha_\mathrm{e}-\alpha_\mathrm{o}) = 2\farcs0$ on average over the
observation, where $(\alpha_\mathrm{o}, \delta_\mathrm{o})$ and
$(\alpha_\mathrm{e}, \delta_\mathrm{e})$ are the original and estimated
coordinates after correction, respectively.  Thus, a total offset
of $\sim$ 2\farcs2 towards the northeast as compared to the original
pointing, i.e., the comet position, was introduced that explains the
fact that the V polarization beam is closer to the comet nucleus as
shown in Fig.~\ref{fig:pointing}.
Since we have not obtained a mapping observation with HIFI, we are not
able to recover the absolute pointing offset from the morphology of the
brightness distribution in the coma.  However, the 1-$\sigma$ pointing
accuracy within the absolute pointing error (APE)  is known for the
epoch of the \chris{} observation, and estimated to be 1\farcs6 after
the warm star tracker correction is taken into account.  The absolute
pointing uncertainty introduces an additional error in the production
rate estimates of about 1\% that has been taken into account in their
absolute standard deviations.

\begin{figure*}
  \centering
  \begin{tikzpicture}
    \begin{axis}[xlabel={$\nu_\mathrm{LSB}$ [GHz]},
      xmin=557.0115, xmax=557.0485,
      xtick={557.01,557.02,557.03,557.04,557.05},
      ymin=0, ymax=1,
      axis x line*=top,
      axis y line=none,
      x dir=reverse,
      minor x tick num=4,
      ]
    \end{axis}
    \begin{axis}[
      spectrum style, name=wbs,
      xmin= -60.868, xmax= -40.868,
      y dir=reverse,
      ymin=-22e-3, ymax=13e-3,
      minor x tick num=1,
      ]
      \draw[thin,black!25] (axis cs:-50.868,\pgfkeysvalueof{/pgfplots/ymin}) --
			(axis cs:-50.868,\pgfkeysvalueof{/pgfplots/ymax});
      \addplot[blue] table [x index=1,y index=2]
	{1342204014_WBS-H_unfold_fluxcal.dat};
      \addplot[red] table [x index=1,y index=2]
	{1342204014_WBS-V_unfold_fluxcal.dat};
      \addplot[black] table [x index=1,y index=2]
	{1342204014_WBS_unfold_fluxcal.dat};
    \end{axis}
    \begin{axis}[xlabel={$\nu_\mathrm{LSB}$ [GHz]},
      at=(wbs.east), anchor=west,
      xmin=556.917, xmax=556.954,
      xtick={556.91,556.92,556.93,556.94,556.95},
      ymin=0, ymax=1,
      axis x line*=top,
      axis y line=none,
      x dir=reverse,
      minor x tick num=4,
      ]
    \end{axis}
    \begin{axis}[
      spectrum style,
      at=(wbs.east), anchor=west,
      yticklabel pos=right,
      ymin=-13e-3, ymax=22e-3,
      minor x tick num=1,
      ]
      \addplot[blue] table [x index=1,y index=2]
	{1342204014_WBS-H_unfold_fluxcal.dat};
      \addplot[red] table [x index=1,y index=2]
	{1342204014_WBS-V_unfold_fluxcal.dat};
      \addplot[black] table [x index=1,y index=2]
	{1342204014_WBS_unfold_fluxcal.dat};
    \end{axis}
  \end{tikzpicture}
  \caption{
    Unfolded level-2 spectra for the horizontal polarization mixer (blue
    line), vertical polarization mixer (red line), and averaged spectrum
    (black line) obtained with the WBS on UT 1.5 September 2010.
    Baselines are fitted with a linear combination of sine functions
    obtained by masking (-1, 1) \si{\kms} windows around the expected
    position of negative (left panel) and positive (right panel)
    phases of the frequency-switched  line rest frequency of the
    \ce{H2O} \trans{} transition.  Note that the scale in the vertical
    axes is the main brightness temperature corrected by the forward and
    beam efficiencies (the vertical axis scale is inverted in the left
    panel).  The upper horizontal axis is the lower sideband frequency,
    and the lower horizontal axis shows the Doppler velocity with
    respect to the nucleus rest frame.
  }
  \label{fig:wbs_unfolded}
\end{figure*}
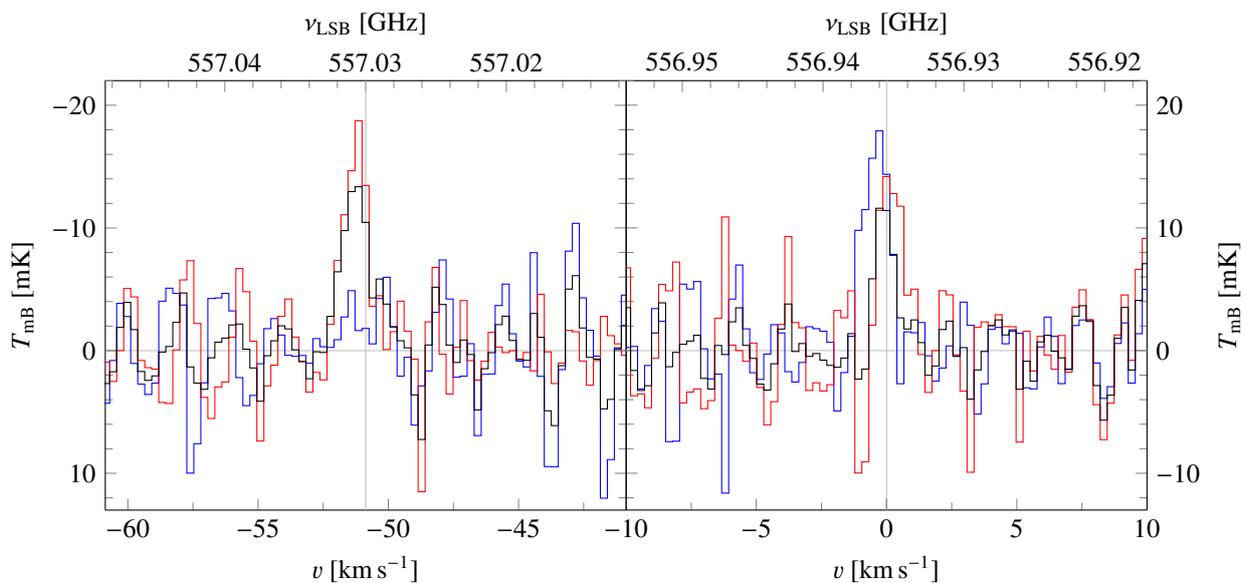

Owing to the small solar phase angle of $\sim$ \SI{11}{\degree}, the
blueshift of the line in the H polarization spectrum might be caused by
preferential outgassing in the direction towards the Sun within
the H aperture.  The smaller line shift in the V spectrum, consistent
with zero offset within uncertainties, suggests a different outgassing
geometry of the material comprised within the V beam.  This indicates
that there may be anisotropies in the coma structure in \chris{}
combined with possible variations in the radial and azimuthal components
of the gas expansion velocity.  Since the line area of the folded V
spectrum is larger than the folded H spectrum and the V beam is pointing
closer to the nucleus position, the difference in the line intensities
in both polarizations may also arise from a higher total column density
averaged within the V beam than in the H beam because of the relative
pointing offset. Using the spherically symmetric ougassing model
described in Sect.~\ref{sec:prodrates}, we expect that the line
intensities should decrease by $\sim$ 1\% and 3\% at
\SIlist{2;5}{\arcsecond} offset, respectively. These values are roughly
consistent with the difference in the line intensity of the folded H and
V spectra, although the deviation in the negative phase of the unfolded
spectra is much larger, probably because of the poor S/N.

Considering that the integrated line of the folded and averaged WBS
spectrum has an S/N of $\sim 4$, it is expected that a line feature is
present in both switching phases at the position of the frequency throw
in the unfolded spectrum if the deteced signal is real.  This aspect
should also be reproducible in both polarizations with an S/N between
2--3.  To test this possibility, we searched for emission features in
the unfolded WBS spectra for H and V polarizations in
Fig.~\ref{fig:wbs_unfolded}.  We found evidence of an emission line
feature in the positive and negative phases of the frequency-switched
line frequency in the V polarization, but a feature is only present in
the positive phase for the H polarization,  although there is no
physical reason to expect that the source has an intrisic polarization
so that the line is present in only one of the polarizations. Generally,
the H polarization has been found to provide better stability and a
lower noise level than the V polarization
\citep[e.g.,][]{2013ApJ...774L...3L}.  Since the S/N in our observation
is limited, it is possible that we do not see a detection in one of the
phases for the H polarization although a signal may be present.

Furthermore, it is likely that a multiple sine wave envelope does not
fully account for the baseline structure around the line emission
feature.  In the frequency switching observing mode without a reference
position, the baseline is not just a combination of sine functions.
Rather, it is fundamentally the difference of the mixer response at one
frequency with that at another frequency.  The fine structure shape of
that response will vary depending on the relative position in the
intermediate frequency (IF), and this can be different in the H and V
polarizations.  Additionally, there is also the possibility the noise
structure around the emission line in the H spectrum is particularly
unfavorable, with several channels at negative noise level, creating a
local drop of the same magnitude as the expected line emission signal.

\subsubsection{\texorpdfstring{\ce{H2O} and \ce{NH3}}{water and ammonia}
production rates}
\label{sec:prodrates}

To compute the production rates we used a spherically symmetric
radiative transfer model that takes into account nonthermal processes
and has previously been applied to analyze \herschel{} data and
ground-based millimeter/submillimeter observations of line emission in
cometary atmospheres \citep[see
e.g.,][]{2010A&A...518L.150H,2010A&A...521L..50D,2011Natur.478..218H,2012A&A...545A...2D,2013A&A...559A..48D}.
The excitation of water molecules was calculated with a code based on
\ratran{} \citep{2000A&A...362..697H,2009AJ....137.4837H}, which
includes collisions between neutrals and electrons and
radiation-trapping effects.  We considered the effects of pumping of the
vibrational excited states by infrared solar radiation.  To compute the
upper limit to the \ce{NH3} production rate, the collisions with
neutrals and electrons were modeled following
\citet{2012A&A...539A..68B}.

We adopted similar parameters as in the models used to derive
production rates from \chris{} observations with  IRAM
\citepalias{2010A&A...518L.149B}.  Since the electron density in the
coma is poorly constrained, an electron density scaling factor of
$\xne = 0.2$ with respect to the standard profile derived from
observations of comet 1P/Halley was used \citepalias[see
e.g.,][]{2007P&SS...55.1058B,2010A&A...518L.149B,2010A&A...518L.150H}.
This value was found to explain the brightness distribution of
the \trans{} water emission line obtained with mapping observations
\citep[e.g.,][]{2007P&SS...55.1058B,2010A&A...518L.150H,2011Natur.478..218H,2012A&A...544L..15B}.
The radial number density profile for each molecule was obtained using
the standard spherically symmetric Haser distribution
\citep{1957BSRSL..43..740H}:
\begin{equation}
  n(r)= \frac{Q}{4\pi r^2v_\mathrm{exp}}\,
  \exp\left(-\frac{r\beta}{v_\mathrm{exp}}\right),
\end{equation}
where $Q$ is the total production rate of the molecule in \si{\mols},
$v_\mathrm{exp}$ is the expansion velocity in the coma, $r$ is the
nucleocentric distance, and $\beta$ denotes the photodissociation rate
including dissociation and ionization effects by the radiation from the
Sun.  The expansion velocity is assumed to be \SI{0.4}{\kms} in comet
\chris{}.  This value is similar to the IRAM observations scaled to 5 AU
and with the expected expansion velocity derived from the half width at
half maximum (HWHM) of a Gaussian fit to the emission line of the
\ce{H2O} transition observed with HIFI.  For low-activity and distant
comets a similar expansion velocity close to \SI{0.5}{\kms} is obtained
from the shapes of the OH line observed at the Nan\c{c}ay radio
telescope \citep{2007A&A...467..729T}.

Generally, the actual outgassing geometry in cometary atmospheres can be
different from the Haser density pattern used in our model.
Nonetheless, the systematic errors introduced by the assumption of a
spherically symmetric distribution are on the same order as the
statistical errors.  Provided that the model uses the same average
parameters, a spherically symmetric outflow provides a similar fit to
the production rate as a nonisotropic distribution consisting of a
combination of several jets with variable velocity, although the shape
of the derived line profile for a nonisotropic outflow may differ
substantially.

Table~\ref{tbl:q} shows the line intensity and rms noise of the
brightness temperature and integrated line area for each spectrometer
that were used to derive the \ce{H2O} production rates and a 3-$\sigma$
upper limit for the \ce{NH3} outgassing rate.  Theoretically, the HRS
and WBS noise levels should be consistent taken into account the square
root of the ratio of their respective resolutions, and the HRS
efficiency, $\eta = 0.81$.  The data have a ratio of the HRS to WBS
noise that is slightly larger than the expected theoretical value.

\begin{table*}
  \caption{Standard deviation of the brightness temperature in the HIFI
    spectra calculated for each spectrometer using the native spectral
    resolution, line intensities, line velocity shifts and retrieved
    \ce{H2O} and \ce{NH3} production rates in comet \christensen{}.}
  \label{tbl:q}
  \centering
  \begin{tabular}{ccc
		  S[table-format = <2.1(2)]
		  S[table-format = +2(2)]
		  S[table-format = <1.1(1)e2]
		  }
    \toprule
    Molecule & Spec. &
    $\sigma_{T_\mathrm{mB}}$\tablefootmark{\emph{a}} &
    ${\int T_\mathrm{mB}\, dv}$ &
    {$\Delta v$\tablefootmark{\emph{b}}} &
    {$Q$\tablefootmark{\emph{c}}}\\
    & & (\si{\milli\K}) & {(\si{\Kms})} & {(\si{\m\per\s})} &
    {(\si{\mols})} \\
    \midrule
    \multirow{2}{*}{\ce{H2O}} & WBS & 1.5 & \wbsho & \wbsv & \wbsqho \\
                              & HRS & 6.3 & \hrsho & \hrsv & \hrsqho \\
    \midrule
    \multirow{2}{*}{\ce{NH3}} & WBS & 1.5 & \wbsn & & \wbsqn \\
                              & HRS & 6.3 & \hrsn & & \hrsqn \\
    \bottomrule
  \end{tabular}
  \tablefoot{
    \tablefoottext{\emph{a}}{Standard deviation computed for
      frequency resolutions of \SI{1.1}{\mega\hertz} and
      \SI{120}{\kilo\hertz} for WBS and HRS.}
    \tablefoottext{\emph{b}}{ The velocity of the line centroids are
      computed using the spectra with the original frequency
      resolution of \SIlist{500;60}{\kilo\hertz} for WBS and HRS.}
    \tablefoottext{\emph{c}}{
    Production rates or 3-$\sigma$ upper limits are derived for a gas
    kinetic temperature of \SI{15}{\K}, expansion velocity of
    \SI{0.4}{\kms}, both values being derived by scaling observations
    obtained at 3.3 AU from the Sun with the IRAM 30-m telescope, and an
    electron density scaling factor of $\xne = 0.2$ \citepalias[see][and
    references therein]{2010A&A...518L.149B}.}
  }
\end{table*}

Since CO is the most abundant species in the atmosphere of \chris{}
\citep[e.g.,][]{2012ApJ...752...15O}, the main collisional excitation
mechanism for \ce{H2O} molecules is collisions with CO molecules.  The
collisional excitation between \ce{H2O} and CO molecules was modeled
using a cross section $\sigma_\ce{CO-H2O} \sim 0.5 \times
\sigma_\ce{H2O-H2O}$, where $\sigma_\ce{H2O-H2O} =
\SI{5e-14}{\cm\squared}$ \citep[see][and references
therein]{1987A&A...181..169B}.  We also investigated a higher collision
rate for \ce{H2O} of \SI{1.5e-13}{\cm\squared}, but with a
$Q_\ce{CO}/Q_\ce{H2O}$ ratio of $\sim$ 6 at 5 AU from the Sun, the
retrieved $Q_\ce{H2O}$ decreases by only 5\%.

Using the standard deviation of the brightness temperature in the WBS
backend, we derive a 3-$\sigma$ upper limit for the line intensity of
the \ce{NH3} line of \SI{\wbsn}{\Kms} over a (-1, 1) \si{\kms} window,
which converts into \SI{14.8}{\Kms} taking into account losses of
hyperfine structure of the line to get 100\% of the line emission.  For
this value, an upper limit of the production rate, $Q_\ce{NH3}
\SI{\wbsqn}{\mols}$, is derived, considering collisions with CO and
using a cross section $\sigma_\ce{CO-NH3} = \SI{3e-14}{\cm\squared}$
(i.e.\ six times larger than $\sigma_\ce{H2O-H2O}$).  From the
3-$\sigma$ upper limit for the line intensity of the HRS \ce{NH3} spectrum
of \SI{\hrsn}{\Kms}, we derive an upper limit to the production rate of
\SI{\hrsqn}{\mols} taking into account losses of hyperfine structure.
Although the collisional cross section between \ce{CO} and \ce{NH3} is
poorly determined, the dominating factor for deriving an \ce{NH3}
production rate of \SI{\wbsqn}{\mols} is radiative self-absorption.

There is a 60\% difference in the production rate calculated in local
thermodynamic equilibrium (LTE), while including only collisions between
water and \ce{NH3}, the production rate $Q_\ce{NH3}$ changes by 10\%.
Therefore, a non-LTE treatment like the escape probability method or a
Monte Carlo radiative transfer approach is required to accurately
compute the \ce{NH3} population levels \citep{2012A&A...539A..68B}.

Compared with the typically observed $Q_\ce{NH3}/Q_\ce{H2O}$ mixing
ratio of about 0.5--1\% in comets at a heliocentric distance of 1 AU
\citep[see e.g.,][and references therein; Biver et al.\ in
preparation]{2012A&A...539A..68B}, the derived upper limit in comet
\chris{} of \SI{\qnh}{\mols} is on the order of the water production
rate, giving a mixing ratio $Q_\ce{NH3}/Q_\ce{H2O} < 0.75$.  However,
because \ce{NH3} is much more volatile than water (with a sublimation
temperature, $T_\mathrm{sub}$, of 78 K to be compared with
$T_\mathrm{sub}$ = 152 K for \ce{H2O}) we could expect an enhanced ratio
$Q_\ce{NH3}/Q_\ce{H2O}$ closer to unity at 5 AU from the Sun.
Nonetheless, a mixing ratio of $\sim$ 1\% cannot be excluded from the
HIFI data.

\subsubsection{Comparison with pre- and post-perihelion observations}

The gas production rate at large heliocentric distances is controlled by
sublimation of highly volatile species like carbon dioxide and carbon
monoxide, which have sublimation temperatures of $\sim$ 70 K.  Although
in most cometary atmospheres \ce{CO2} has been found to dominate
\ce{CO} with a wide variety of mixing ratios
\citep[e.g.,][]{2010ApJ...717L..66O,2012ApJ...752...15O}, the activity
of \chris{} was mostly driven by CO emission from the measurements
obtained around 3.1--3.7 AU.  Scaling the CO production rate obtained
from post-perihelion measurements of the CO (2--1) line at IRAM by
$\rh^{-2}$, as measured in comet \object{C/1995 O1 (Hale-Bopp)}
\citep{1997EM&P...78....5B}, we estimate a mixing ratio
$Q_\ce{CO}/Q_\ce{H2O}$ of 6.5 at 5 AU.  These IRAM observations were
obtained almost simultaneously with the PACS and SPIRE measurements from
November 2009.  The CO production rate measured by IRAM is about a
factor of two higher than the value derived by the \akari{} infrared
observations at $\rh$ = 3.13 AU \citep{2012ApJ...752...15O}.  This might
be explained by the difference in the field of view of the observations
and uncertainties in the coma distribution and excitation parameters
\citep[e.g.,][]{2012ApJ...752...15O}.  Additionally, infrared detections
of the hypervolatiles CO, \ce{CH4} and \ce{C2H6} were obtained in comet
\chris{} using the Cryogenic Infrared Echelle Spectrograph (CRIRES) on
the Very Large Telescope (VLT) at heliocentric distances of 3.25 AU and
4.03 AU, as initially reported in
\citet{2010DPS....42.2821B,2013DPS....4541318B}.

\begin{figure}
  \centering
  \begin{tikzpicture}
    \begin{semilogyaxis}[
	  only marks,
	  xlabel={$\rh$ [AU]},
	  ylabel={$Q$ [\si{\mols}]},
	  ymin=1e27,xmin=3, xmax=5.5,
    ]
      \addplot[green,mark=triangle,error bars/.cd,y dir=both,y explicit]
	coordinates {
	  (3.66, 299.3e26) +- (0., 30.0e26)
	  (3.13, 197.7e26) +- (0., 19.9e26)};
      \addplot[green,mark=diamond*,error bars/.cd,y dir=both,y explicit]
	coordinates {(3.20, 3.9e28) +- (0., 3e27)
		    (3.32, 3.0e28) +- (0., 3e27)};
      \addplot[red,mark=square*,error bars/.cd,y dir=plus,
	y fixed relative=0.58]
	coordinates {
	  (3.14, 1.5e28)
	  (4.45, 2.95e27)};
      \addplot[red,mark=square*,error bars/.cd,y dir=minus,
	y fixed relative=0.369]
	coordinates {
	  (3.14, 1.5e28)
	  (4.45, 2.95e27)};
      \addplot[red,mark=triangle,error bars/.cd,y dir=both,y explicit]
	coordinates {
	  (3.66, 84.59e26) +- (0., 8.46e26)
	  (3.13, 84.66e26) +- (0., 8.47e26)};
      \addplot[red,mark=triangle*,mark options={rotate=180},error
	bars/.cd,y dir=both,y explicit]
	coordinates {(3.13, 76.38e26) +- (0., 60.3e26)};
      \addplot[blue,mark=*,error bars/.cd,y dir=both,y explicit]
	coordinates {(5., 2.e27) +- (0., 5e26)};
      \addplot[blue,mark=*]
	coordinates {(3.35, 1.4e28) (3.34, 4.e28)};
      \draw[blue,->] (axis cs:3.35,1.4e28) -- (axis cs:3.35,1e28);
      \draw[blue,->] (axis cs:3.34,4e28) -- (axis cs:3.34,2.86e28);
      \addplot[blue,mark=triangle,error bars/.cd,y dir=both,y explicit]
	coordinates {
	  (3.66, 82.96e26) +- (0., 8.55e26)
	  (3.13, 201.0e26) +- (0., 20.3e26)};
      \addplot[blue,mark=pentagon,error bars/.cd,
	x dir=both,x explicit,y dir=both,y explicit]
	coordinates {(3.27, 4.2e28) +- (0.16, 1e28)}; 
      \addplot[mark=none,smooth] table [x index=1,y index=7]
	{Sub_Ch.txt};
    \end{semilogyaxis}
  \end{tikzpicture}
  \caption{
    \ce{H2O} (blue data points), \ce{CO2} (red data points) and CO (green
    data points) production rates in comet \christensen{} as a function of
    heliocentric distance with 1-$\sigma$ uncertainties.  Open symbols
    show pre-perihelion production rates and filled symbols are
    post-perihelion rates. Circles represent detections by \herschel{}
    \citepalias[][and this work]{2010A&A...518L.149B}; triangles are
    production rates from \akari{} \citep{2012ApJ...752...15O}; squares
    are rates measured by \spitzer{} \citep{2013Icar..226..777R}; the
    inverted triangle is the \ce{CO2} production rate computed from from
    the $N_\ce{CO2}/N_\ce{H2O}$ ratio estimated from the \ce{OI}
    forbidden lines measured at the Apache Point Observatory 3.5-m
    telescope \citep{2012Icar..220..277M};  diamonds are production
    rates obtained by the IRAM 30-m telescope, and the pentagon is the
    $Q_\ce{H2O}$ derived from OH observations at Nan\c{c}ay
    \citepalias{2010A&A...518L.149B}.  The upper limits to the water
    production rate at about 3.3 AU were obtained by PACS and SPIRE
    spectroscopy \citepalias{2010A&A...518L.149B}.  The solid line is
    the water production rate by sublimation at the subsolar point on
    the nucleus.
  }
  \label{fig:prodrates}
\end{figure}
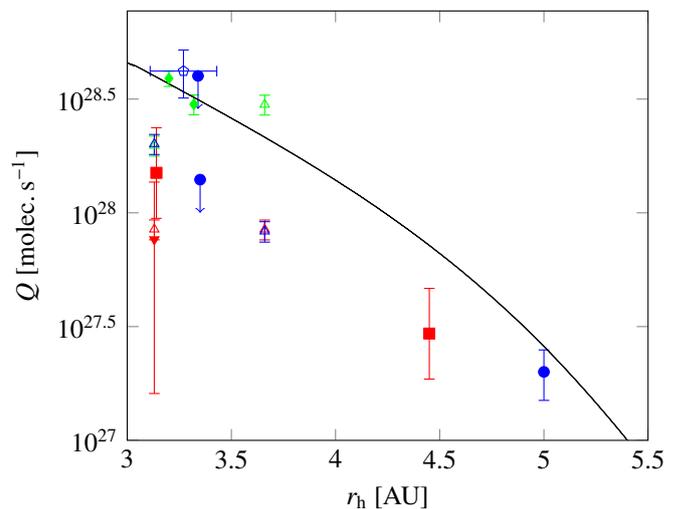

Figure~\ref{fig:prodrates} shows the \ce{H2O}, \ce{CO2} and CO
production rates as a function of heliocentric distance as measured by
various instruments.  The water production rate measured by HIFI and the
production rate at the subsolar point on the nucleus shown by the solid
curve in Fig.~\ref{fig:prodrates} agree but this similitude does not
require water emission to come entirely from the nucleus. The solid
curve represents the maximum value of water production rate because it
is assumed that there is no dust on the surface and the thermal inertia
of the surface ice is very low. The absorbed solar energy goes into
sublimation, thermal re-radiation, and diffusion of heat into the
interior of the nucleus in our model.

However, this simple model does not include release of trapped CO
molecules during crystallization of amorphous water ice, gas diffusion,
and re-condensation of gas in deeper, cooler layers, or
sublimation of icy-dust grains in the coma.  More detailed studies of
the chemically differentiated nucleus and the CO sublimation are needed
but they are beyond the scope of the current paper.  Nonetheless, the
\ce{H2O} production rate profile shown in Fig.~\ref{fig:prodrates}
satisfies the observational data to a first approximation, as
described by \citet{2012LPICo1667.6192S}.  The profile was retrieved
assuming a size of the active area of \SI{48}{\km^2}, that is, covering
about 4\% of the surface considering a spherical nucleus with a radius
of \SI{10}{\km}.

It is very likely that part of the water molecules in the coma of comet
\chris{} at large distance from the Sun are produced by the sublimation
of small icy grains \citepalias[see discussion
in][]{2010A&A...518L.149B}.  Evidence for sublimation from water ice
grains have been observed previously by \herschel{} in other comets, for
instance, \object{C/2009 P1 (Garradd)}
\citep{2012A&A...544L..15B,2012LPICo1667.6330B,2014A&A...562A...5B} and
\object{103P/Hartley 2} \citep{2011Natur.478..218H,
2011ApJ...734L...1M}, and at large heliocentric distance in comet C/1995
O1 (Hale–Bopp) \citep{1997Icar..127..238D}.  However, there is also
evidence of nuclear emission in \chris{}; the water emission line in the
averaged WBS spectrum is blueshifted by \SI{\wbsv}{\m\per\s}, consistent
with outgassing from the nucleus in the direction towards the Sun.
Nevertheless,  we cannot exclude the possibility of \ce{H2O} sublimation
from small ice-bearing grains outflowing from the nucleus surface in
comet \chris{}.

\begin{figure*}
  \centering
  \begin{tikzpicture}
    \begin{axis}[map style,name=red1]
      \addplot graphics
         [xmin=-30,xmax=30,ymin=-30,ymax=30]
         {1342186621_red};
      \draw[->,thick] (axis cs:-9.766,-2.153) --
        (axis cs:-19.531,-4.306);
      \draw[->,dashed,thick] (axis cs:-3.907,-9.205) --
	(axis cs:-7.815,-18.410);
      \node at (axis cs:-24.414,-5.382) {\sun};
      \addplot[only marks,mark=+,color=white] coordinates {(0,0)};
    \end{axis}
    \begin{axis}[map style, name=blue1,
      at=(red1.west),anchor=east
      ]
      \addplot graphics
           [xmin=-30,xmax=30,ymin=-30,ymax=30]
           {1342186621_blue};
      \draw[->,thick] (axis cs:-9.766,-2.153) --
	(axis cs:-19.531,-4.306);
      \draw[->,dashed,thick] (axis cs:-3.907,-9.205) --
	(axis cs:-7.815,-18.410);
      \node at (axis cs:-24.414,-5.382) {\sun};
      \addplot[only marks,mark=+,color=white] coordinates {(0,0)};
    \end{axis}
    \begin{axis}[map style,name=red,
      at=(red1.south),anchor=north
      ]
      \addplot graphics
	[xmin=-30,xmax=30,ymin=-30,ymax=30]
	{1342203478_red};
      \draw[->,thick] (axis cs:-9.964,-0.851) --
	(axis cs:-19.927,-1.702);
      \draw[->,dashed,thick] (axis cs:-6.820,-7.314) --
	(axis cs:-13.640,-14.627);
      \node at (axis cs:-24.909,-2.127) {\sun};
      \addplot[only marks,mark=+,color=white] coordinates {(0,0)};
    \end{axis}
    \begin{axis}[map style,name=blue,
	at=(red.west),anchor=east
	]
      \addplot graphics
	[xmin=-30,xmax=30,ymin=-30,ymax=30]
	{1342203478_blue};
      \draw[->,thick] (axis cs:-9.964,-0.851) --
	(axis cs:-19.927,-1.702);
      \draw[->,dashed,thick] (axis cs:-6.820,-7.314) --
	(axis cs:-13.640,-14.627);
      \node at (axis cs:-24.909,-2.127) {\sun};
      \addplot[only marks,mark=+,color=white] coordinates {(0,0)};
    \end{axis}
  \end{tikzpicture}
  \caption{
    Maps at \SI{70}{\um} (blue band; \emph{left column}) and
    \SI{160}{\um} (red band; \emph{right column}) of comet
    \christensen{} observed with PACS on UT 1.5 November 2009
    \citepalias[\emph{upper row}, see][]{2010A&A...518L.149B} and UT
    26.5 August 2010 (\emph{bottom row}) centered on the peak position
    indicated by the white cross.  The pixel sizes are 3.2\arcsec{} in
    the blue map and 6.4\arcsec{} in the red map for the November 2009
    observations, and 1\arcsec{} in the blue map and 2\arcsec{} in the
    red map for the August 2010 data.  Contour levels have a step of 0.1
    on logarithmic scale in all the images. The position of the nucleus
    at the mid-time of the observations according to the HORIZONS
    ephemeris is shown by the yellow circle.  The projected directions
    toward the Sun and velocity vector of the comet are indicated by the
    solid and dashed arrows.
    }
  \label{fig:pacs}
\end{figure*}
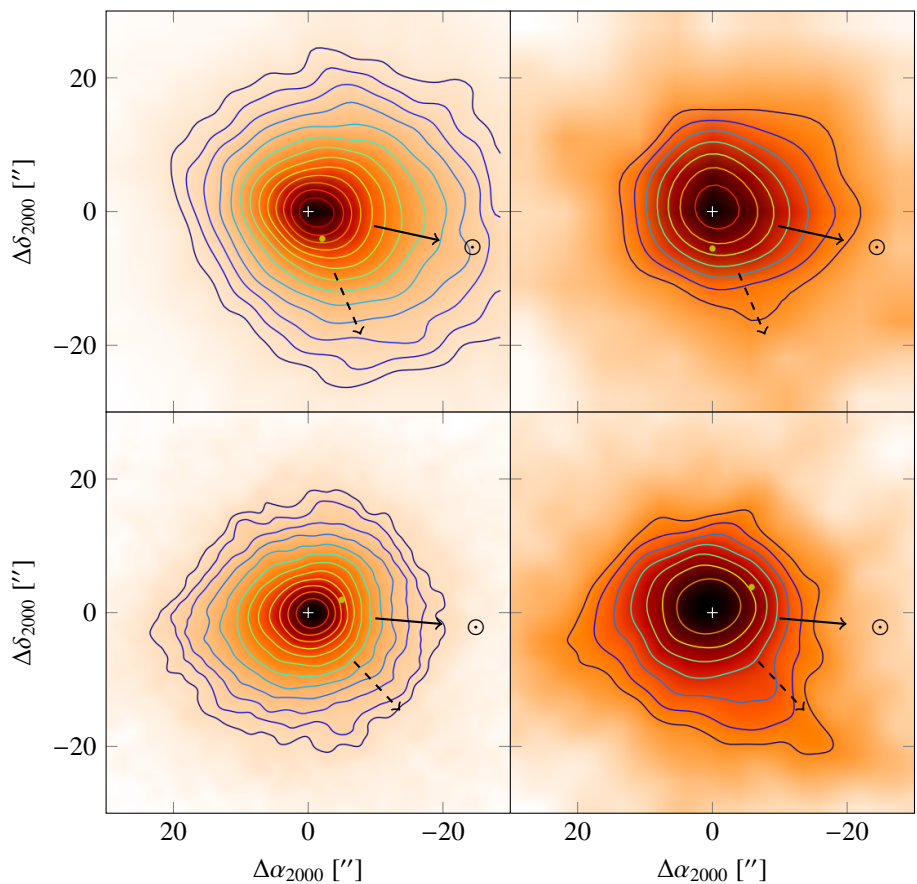

\subsection{PACS data analysis}
\label{sec:resultspacs}

We analyzed the PACS photometric observations in the same way as
described in \citetalias{2010A&A...518L.149B}.  The proper motion of the
comet was relatively fast at the time of the observations, of about
\SI{10}{\arcsec\per\hour}, and the total integration time was $\sim$
\SIrange{10}{20}{\minute} for each scanning direction in the PACS
observations.  Because of this, the Astronomical Observation Requests
(AORs) for these observations identified comet \chris{} as a moving
solar system object and the nonsidereal tracking was activated by the
mission planning system that lead to a continuous tracking of the motion
of the comet.  The HIPE pipeline that has been used to reduce the PACS
observations with this special tracking mode brings all the individual
frames together for each individual scan, correcting for the observed
proper motion in this observing mode.  For each scanning direction,
individual maps were projected onto the same World Coordinate System
(WCS) centered on the comet nucleus by shifting the data arrays onto a
uniform grid with multivariate interpolation using a fixed Gaussian
kernel.  We subtracted the background emission and averaged the maps
with orthogonal scanning direction, correcting for the proper motion of
the target during the observations for each of the two epochs \citep[see
e.g.,][]{2012A&A...541A..92S} using the ephemeris computed by the
HORIZONS system \citep{1996DPS....28.2504G}.  The position of the
nucleus at mid-time of the observations was used to center each of the
images.

Figure~\ref{fig:pacs} shows the blue and red maps obtained with PACS in
November 2009---presented in \citetalias{2010A&A...518L.149B}---and
August 2010, centered on the peak position of the images after merging
the observation pairs with orthogonal scan directions for each channel.
The nucleus position at mid-time of the observations is shown by the
yellow circles. PACS photometer observations obtained in the scan
mapping mode can be affected by offsets due to the 1-$\sigma$ APE of the
telescope of about \SIrange{1}{2}{\arcsecond} at the time the \chris{}
observations were acquired \citep[see][and references
therein]{2010A&A...518L...1P,2013ExA...tmp...47V}. This value can
explain the offset between the nucleus position and the estimated
position of the peak shown in Fig.~\ref{fig:pacs}.  On the other hand,
the nominal relative pointing error (RPE) measured once the telescope is
commanded to the target position using short observing intervals of
$\sim$ \SI{1}{\minute} during the performance verification (PV) phase
has been found to lie in the range 0\farcs2--0\farcs3 for solar system
objects within 68\% confidence level.  This error can generally be
neglected compared with the FWHM beam sizes of \SIlist{5;11}{\arcsecond}
for the \SIlist{70;160}{\um} PACS scan-mapping observations.

\subsubsection{Dust coma distribution and radial profiles}

The surface brightness of the \SI{70}{\um} image has a broader
distribution than the \SI{160}{\um} image in both epochs, although the
difference is significantly smaller in the August 2010 observations (see
lower panel in Fig.~\ref{fig:pacs}).  The coma distribution clearly
broadens in the direction towards the Sun in the first observations,
possibly due to outflow of large dust grains.  In the more recent PACS
maps, the red band image is elongated towards the southwest close to the
direction of the velocity vector of the comet, and the blue image is
roughly symmetric.  Note that the solar phase angle was \SI{11}{\degree}
at the time of the latter observations, so the observing geometry is
not the most appropriate to see an extended dust tail towards the
antisolar direction.  Arc-like asymmetries and spirals are observed in
the \SI{4.5}{\um} \spitzer{} images of \chris{} when the comet was at
3.1 AU from the Sun. However, these can be explained in part by gas
structures \citep{2013Icar..226..777R}.

\begin{figure*}
  \centering
  \begin{tikzpicture}
    \begin{loglogaxis}[profile style, name=2009,
	ymin = 1e-5, ymax = 5e-2,
	]
      \addplot[blue,mark size=.5pt] table [x index=0,y index=1]
       {1342186621_blue_60_prof.dat};
      \addplot[red,mark size=.5pt] table [x index=0,y index=1]
       {1342186621_red_60_prof.dat};
      \addplot[green,mark=x, error bars/.cd,y explicit,y dir=both]
       table [x index=0,y index=1,y error index=2]
       {1342186621_blue_1.0_60_prof.dat};
      \addplot[mark=none,green,smooth] table [x index=0,y index=1]
       {1342186621_blue_1.0_40_prof_fit.dat};
      \addplot[green,mark=o, error bars/.cd,y explicit,y dir=both]
       table [x index=0,y index=1,y error index=2]
       {1342186621_red_2.0_60_prof.dat};
      \addplot[mark=none,green,smooth] table [x index=0,y index=1]
       {1342186621_red_4.0_40_prof_fit.dat};
    \end{loglogaxis}
    \begin{loglogaxis}[profile style,
	at=(2009.east), anchor=west,
	yticklabel=\empty, ylabel=\empty,
	ymin = 1e-5, ymax = 5e-2,
	]
      \addplot[blue,mark size=.5pt] table [x index=0,y index=1]
	{1342203478_blue_60_prof.dat};
      \addplot[red,mark size=.5pt] table [x index=0,y index=1]
	{1342203478_red_60_prof.dat};
      \addplot[green,mark=x, error bars/.cd,y explicit,y dir=both]
	table [x index=0,y index=1,y error index=2]
	{1342203478_blue_0.5_60_prof.dat};
      \addplot[mark=none,green,smooth] table [x index=0,y index=1]
	{1342203478_blue_1.0_40_prof_fit.dat};
      \addplot[green,mark=o, error bars/.cd,y explicit,y dir=both]
	table [x index=0,y index=1,y error index=2]
	{1342203478_red_1.0_60_prof.dat};
      \addplot[mark=none,green,smooth] table [x index=0,y index=1]
	{1342203478_red_1.0_40_prof_fit.dat};
    \end{loglogaxis}
  \end{tikzpicture}
  \caption{
    \emph{Left panel}: surface brightness in comet \christensen{} as a
    function of distance from the position of the center of the peak
    measured with PACS on UT 1.8 November 2009 (shown by the blue and
    red dots for the \SIlist{70;160}{\um} bands with a pixel size of
    3\farcs2 and 6\farcs4).  \emph{Right panel}: surface brightness
    radial profiles observed on UT 26.5 August 2010 (blue and red dots
    denote the \SIlist{70;160}{\um} bands with projected pixel size of
    \SIlist{1;2}{\arcsecond}).  The green crosses and circles are
    resampled profiles with 1\arcsec{} and 2\arcsec{} bins for the blue
    and red band images from November 2009, and 0\farcs5 and 1\arcsec{}
    bins for the blue and red band images from August 2010.  Error bars
    are 1-$\sigma$ statistical uncertainties.
    }
  \label{fig:profiles}
\end{figure*}
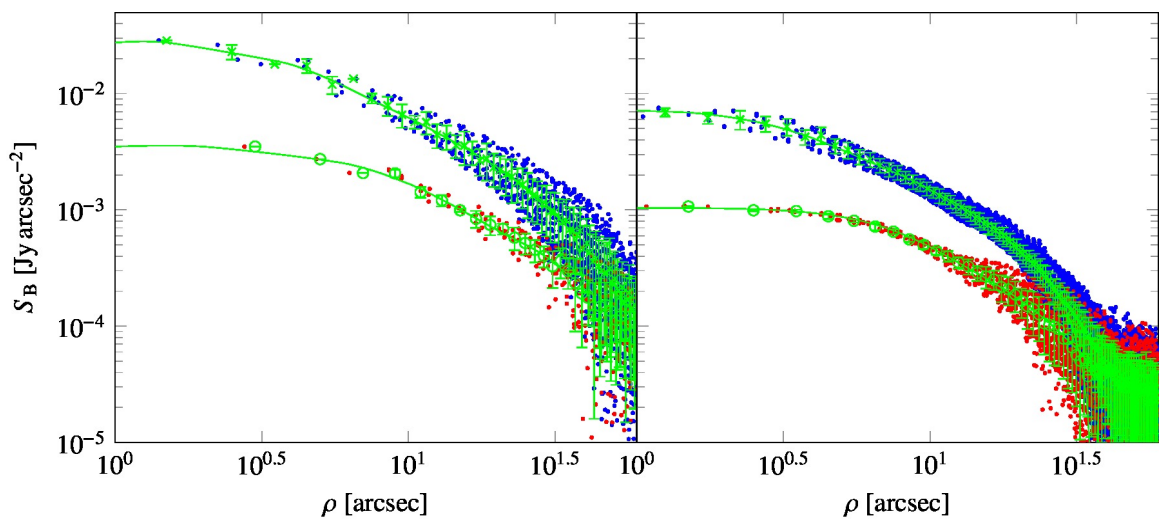

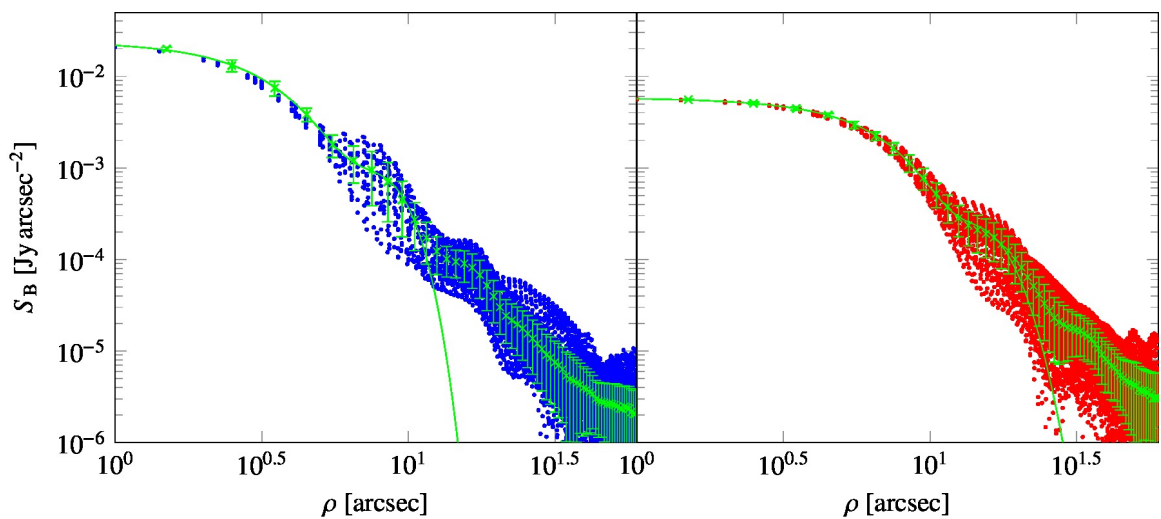
\begin{figure*}
  \centering
  \begin{tikzpicture}
    \begin{loglogaxis}[profile style, name=blue,
      ymin=1e-6, ymax=5e-2,
      ]
      \addplot[blue,mark size=.5pt] table [x index=0,y index=1]
	{PSF_blue_0_prof.dat};
      \addplot[green,mark=x, error bars/.cd,y explicit,y dir=both]
	table [x index=0,y index=1,y error index=2]
	{PSF_blue_1.0_prof.dat};
      \addplot[mark=none,green,smooth] table [x index=0,y index=1]
	{PSF_blue_gauss.dat};
    \end{loglogaxis}
    \begin{loglogaxis}[profile style,
	ymin=1e-6, ymax=5e-2,
	yticklabel=\empty, ylabel=\empty,
	at=(blue.east), anchor=west]
      \addplot[red,mark size=.5pt] table [x index=0,y index=1]
	{PSF_red_0_prof.dat};
      \addplot[green,mark=x, error bars/.cd,y explicit,y dir=both]
	table [x index=0,y index=1,y error index=2]
	{PSF_red_1.0_prof.dat};
      \addplot[mark=none,green,smooth] table [x index=0,y index=1]
	{PSF_red_gauss.dat};
    \end{loglogaxis}
  \end{tikzpicture}
  \caption{
      Radial PSF profiles from PACS observations of the main-belt
      asteroid Vesta in the blue (\SI{70}{\um}; \emph{left panel}) and
      red (\SI{160}{\um}; \emph{right panel}) channels. The green
      crosses are the resampled data using a 1\arcsec{} bin with
      1-$\sigma$ statistical uncertainties. Solid lines are fits to the
      resampled PSF profiles using a sum of two Gaussian functions.
    }
  \label{fig:psf}
\end{figure*}

The radial profiles of surface brightness with respect to the peak
position in the blue and red bands are shown in Fig.~\ref{fig:profiles}.
We fitted the resampled radial profiles by convolving the function
$\rho^{-x}$ with a fit to the PSFs of \object{Vesta} consisting of two
Gaussian functions.  Figure~\ref{fig:psf} shows the radial profiles of
surface brightness in the reference PSFs from Vesta data with respect to
the center of the peak position in the blue and red bands, and fits to
the profiles.  We find that the observed surface brightness profiles of
\chris{} can be reproduced by exponents in the range $x$=0.8--0.9 in the
observations from August 2010, while a value of $x$=1.0--1.2 better fits
the profiles from November 2009 \citepalias{2010A&A...518L.149B}.  We do
not find evidence for a substantial contribution from the nucleus
thermal emission to the flux of the central pixels.  A comparison with
observations at optical wavelengths during the same period would provide
further constraints on the dust model described in
Section~\ref{dustmodel}, but these observations are not available in the
literature.

\subsubsection{Dust size distribution and production rates}
\label{dustmodel}

\begin{table}
\caption{Flux densities from PACS observations of comet \christensen.}
\label{tbl:obsp}
  \centering
  \begin{tabular}{c
		  S[table-format = 2.1(2)]
		  S[table-format = 1.2(3)]}
    \toprule
    Date & \multicolumn{2}{c}{Flux at
    photocenter\tablefootmark{\emph{a}}}\\
    & {\SI{70}{\micro\m}} & {\SI{160}{\micro\m}} \\
    (yyyy-mm-dd) & {(mJy/pxl)} & {(mJy/pxl)} \\
    \midrule
    2009-11-01\rlap{\tablefootmark{\emph{b}}} & 37.3(2) & 5.0(3) \\
    2010-08-26      & 8.6(2)  & 1.27(6) \\
    \bottomrule
  \end{tabular}
\tablefoot{
  \tablefoottext{\emph{a}}{ The pixel size is 1\arcsec{} for both
  channels. The quoted error bar does not take into account a 5\%
  uncertainty in the flux calibration.}
  \tablefoottext{\emph{b}}{Data from \citetalias{2010A&A...518L.149B}.}
}
\end{table}

To determine the dust production rate $Q_\mathrm{dust}$, we
compared the flux densities measured on the brightest pixels of the blue
and red PACS maps with those expected from a model of dust thermal
emission.  Table~\ref{tbl:obsp} shows the measured flux densities at
photocenter by PACS above the background noise in the two observing
epochs.

The model used for this study is the same as that applied to the PACS
data of comet \chris{} obtained in 2009 at 3.35 AU from the Sun
\citepalias{2010A&A...518L.149B}, comet \object{C/2012 S1 (ISON)}
\citep{2013A&A...560A.101O} and of the centaurs \object{2060 Chiron} and
\object{10199 Chariklo} \citep{2013A&A...555A..15F}. The basic
principles of this model are given in \citet{1990ApJ...365..738J}.
Absorption cross-sections calculated with the Mie theory were used to
compute the temperature of the grains, solving the equation of radiative
equilibrium, and their thermal emission. Complex refractive indices of
amorphous carbon and olivine (Mg:Fe = 50:50)
\citep{edoh83,1995A&A...300..503D} were taken as broadly representative
of cometary dust.

We considered a differential dust production $Q_\mathrm{dust}(a)$ as a
function of grain radius $a$, described by the size index $\alpha$. The
size-dependent grain velocities $v_\mathrm{dust}(a)$, as well as the
maximum grain radius $a_\mathrm{max}$, were computed following
\citet{1997Icar..127..319C}.  We assumed a nucleus bulk density of
\SI{500}{\kg\m^{-3}}, consistent with observational constraints
\citep{2007Icar..187..306D,2007Icar..191S.176R,2013Icar..222..550T} and
the same value for the dust density consistent with the average density
of the ``fluffy'' dust particles found in the coma of comet
\object{81P/Wild 2} \citep{2012ApJ...744...18N}.

The maximum grain size and dust velocities critically depend on the
nucleus size and the gas production rate at the surface, which are both
poorly constrained in \chris{}. For the nominal value of the nucleus
size, our model assumes a nucleus radius of $R_\mathrm{N} =
\SI{10}{\km}$, as in \citetalias{2010A&A...518L.149B}.

\begin{table*}
\caption{Dust model parameters and dust production rates of comet
\christensen{} derived from PACS \SIlist{70;160}{\micro\m} data for
carbon and olivine grains.}
\label{tbl:qdust}
\centering
  \begin{tabular}{cc
		  S[table-format=1e2]
		  SccS
		  S[table-format=4]
		  S[table-format=4]
		  S[table-format=4]
		  S[table-format=4]}
  \toprule
  Date & $\rh$ & {$Q_\ce{CO}$} & {$a_\mathrm{max}$} &
  \multicolumn{2}{c}{$v_\mathrm{dust}$\tablefootmark{\emph{a}}} & {$\alpha$} &
  \multicolumn{2}{c}{$Q_\mathrm{carbon}$\tablefootmark{\emph{b}}} &
  \multicolumn{2}{c}{$Q_\mathrm{olivine}$\tablefootmark{\emph{b}}} \\
  & & & & $a_\mathrm{min}$ & $a_\mathrm{max}$ & & \SI{70}{\micro\m} & \SI{160}{\micro\m} & \SI{70}{\micro\m} & \SI{160}{\micro\m} \\
  (yyyy-mm-dd) & (AU) & {(\si{\mols})} & {(\si{\mm})} & (\si{\m\per\s}) & (\si{\m\per\s}) & &
  {(\si{\kgs})} & {(\si{\kgs})} & {(\si{\kgs})} & {(\si{\kgs})} \\
  \midrule
             &      &      &                              & & & -3   & 1520 & 1092 & 923 & 1190 \\
  2009-11-08 & 3.35 & 3e28 & 0.59\tablefootmark{\emph{c}} & 5.7 & 174 & -3.5 & 923  & 884  & 676 & 965  \\
             &      &      &                              & & & -4   & 650  & 900  & 606 & 1220 \\
  \midrule
             &      &      &      & & & -3   & 89 & 56  & 63 & 76  \\
  2010-08-26 & 4.96 & 4e27 & 0.07 & 6.3 & 106\tablefootmark{\emph{d}} & -3.5 & 82 & 68  & 70 & 102 \\
             &      &      &      & & & -4   & 98 & 112 & 96 & 182 \\
  \midrule
             &      &      &      & & & -3   & 172 & 102 & 113 & 109 \\
  2010-08-26 & 4.96 & 8e27 & 0.14 & 6.3 & 131\tablefootmark{\emph{d}} & -3.5 & 139 & 104 & 109 & 130 \\
             &      &      &      & & & -4   & 140 & 153 & 133 & 221 \\
  \bottomrule
  \end{tabular}
\tablefoot{
  \tablefoottext{\emph{a}}{Dust velocities for sizes
  $a_\mathrm{min}$ and $a_\mathrm{max}$, with $a_\mathrm{min}$ =
  \SI{0.1}{\micro\m}.}
  \tablefoottext{\emph{b}}{Dust production rate derived assuming
    $a_\mathrm{min} = \SI{0.1}{\micro\m}$.}
  \tablefoottext{\emph{c}}{In \citetalias{2010A&A...518L.149B}, the
  maximum size computed by our dust model was multiplied by 1.5 to
  better match state-of-the-art hydrodynamic simulations
  \citep[see][and references therein]{2005Icar..176..192C}.}
  \tablefoottext{\emph{d}}{The velocity of 1-\si{\um} sized particles is
  \SIlist{45;60}{\m\per\s} for $Q_{\rm CO}$ = \SIlist{4e27;8e27}{\mols},
  respectively.}
}
\end{table*}

Measurements of the CO and \ce{CO2} production rates in the heliocentric
range $\rh$ = 3.1--3.7 AU (Fig. 4) show that the distant activity of
comet \chris{} was CO-dominated, with a $Q_\ce{CO2}/Q_\ce{CO}$
production rate ratio of typically 0.3--0.4
\citep{2012ApJ...752...15O,2012Icar..220..277M}. From the intensity of
the emission observed in the wide 4.5-\si{\micro\m} \spitzer{} filters
on 9 July 2010 at 4.45 AU post-perihelion \citep{2013Icar..226..777R},
which mixes the contributions of fluorescence emission from \ce{CO2} and
CO, we estimate the CO production rate to be $\sim \SI{8e27}{\mols}$ at
that time. On the other hand, the extrapolation of the heliocentric
variation of $Q_\ce{CO}$ from \SIrange{3.66}{4.45}{AU} measured by
\akari{} and \spitzer{} \citep{2012ApJ...752...15O,2013Icar..226..777R},
to the heliocentric distance ($r_h$ = 4.96 AU), corresponding to the
date of the PACS measurements from 26 August 2010, yields $Q_\ce{CO} =
\SI{4e27}{\mols}$.  Therefore, we considered a CO production rate of
\SIlist{4e27;8e27}{\mols} in our modeling, referred to as low- and
high-activity cases.

Table~\ref{tbl:qdust} presents the maximum dust size and the range of
dust velocities computed by our dust model.  Results for 8 November 2009
\citepalias{2010A&A...518L.149B} and 26 August 2010 are presented in the
table. In September 2009 at $\rh$ = 3.35 AU, the CO production rate was
measured to be $Q_\ce{CO} = \SI{3e28}{\mols}$ with the IRAM 30-m
telescope \citepalias{2010A&A...518L.149B}, and this value was used to
calculate the dust parameters (Table~\ref{tbl:qdust}). We considered two
different CO production rates for the August 2010 observations.

Dust production rates derived from the 70-\si{\micro\m} and
160-\si{\micro\m} maps of 26 August 2010 agree within a factor of 1.7 or
lower, depending on the size index (Table~\ref{tbl:qdust}). For the
high-activity case and olivine grains, the best agreement is found for a
size index $\alpha \sim -3.3$, while for carbon grains the best fit is
for $\alpha \sim -4$. For the low-activity case, the best fit is for
$\alpha > -3$ and $\alpha \sim -4$ for carbon and olivine grains. In
conclusion, the PACS data of August 2010 suggest a size index between
\numlist{-4;-3}, consistent with the value $\alpha =
-3.6^{+0.25}_{-0.8}$ derived in \citetalias{2010A&A...518L.149B} from
the August 2009 data.  For the size indexes that explain the fluxes at
the photocenters of the \SIlist{70;160}{\micro\m} maps, we infer dust
production rates in the range \SIrange{70}{110}{\kg\per\s} at 4.96 AU
from the Sun (Table~\ref{tbl:qdust}). This corresponds to a dust-to-gas
production rate ratio of 0.3--0.4. The dust production rate derived in
August 2010 is one order of magnitude lower than in November 2009,
similar to the CO production rate difference in the same period
\citepalias[Table~\ref{tbl:qdust},][]{2010A&A...518L.149B}, suggesting
that the dust-to-gas production rate ratio remained approximately
constant from 3.35 AU to 4.96 AU. When considering a nucleus radius of
$R_\mathrm{N}$ = 5 km instead of 10 km, the inferred dust production
rates and dust-to-gas ratios are typically twice as high, while the
dust-to-gas production rate ratio remains constant over those
heliocentric distances.

\section{Discussion}
\label{sec:discuss}

We studied the gas and dust activity in the long-period comet
\christensen{} using remote-sensing observations obtained with the
\herschel{} Space Observatory at 5.0 AU from the Sun.  A tentative
detection of the ortho-\ce{H2O} ground-state transition is observed in
the HIFI spectra with 4-$\sigma$ significance.  Even though the \ce{H2O}
line is only marginally detected, the derived \ce{H2O} production rate
and line shape and velocity shift are consistent with the emission
feature having a cometary origin.  We derive a water production rate of
\SI{\qho}{\mols} using a spherically symmetric radiative transfer model.
Concurrently, we aimed to detect the ground-state rotational transitions
of ortho-\ce{NH3} in the USB of HIFI's band 1b, from which a 3-$\sigma$
upper limit for the ammonia production rate of \SI{\qnh}{\mols} is
derived, corresponding to a mixing ratio $Q_\ce{NH3}/Q_\ce{H2O} < 0.75$.

Water sublimates fully at heliocentric distances \textless{} 2.5 AU and
is responsible for the activity of cometary nuclei at these distances
form the Sun, while its sublimation becomes inefficient at larger
distances where the nucleus surface temperature may be lower than the
\ce{H2O} sublimation temperature \citep{2000cssr.conf.....C}.  Thus,
outgassing activity at heliocentric distances \textgreater{} 3 AU is
mostly driven by molecular species more volatile than water such as
carbon monoxide.  To date, the largest heliocentric distance at which OH
emission---which is regarded as a proxy for \ce{H2O}---has been detected
is \SI{4.7}{AU}, in pre-perihelion observations of the exceptionally
active comet C/1995 O1 (Hale–Bopp) with the Nan\c{c}ay radio telescope
\citep{1997EM&P...78...37C}.

Whether the observed water emission in \chris{} originates directly in
the nucleus or it is produced by icy grains in the coma can be partly
assessed from the shape of the emission line.  The characteristic
blueshift of the water line detected by the H polarization WBS spectrum
is indicative of preferential emission from the daytime hemisphere in
the direction towards the Sun, taking into account the small solar phase
angle of $\sim$ \SI{11}{\degree} at the time of the observations.  The
smaller blueshift of the line in the V polarization WBS spectrum
suggests that the material in the coma has a nonspherical outgassing
geometry with possibly variable gas velocity.  However, from the HIFI
observations we cannot exclude the possibility that water sublimation
occurs in small icy grains outflowing at a speed close to \SI{0.2}{\kms}
from the nucleus surface.

Production rates of several species measured with ground-based and space
observatories at different heliocentric distances were compared with the
\ce{H2O} production rate derived from the \herschel{} observations.  The
gas production rate in \chris{} is driven by sublimation of
hyper-volatile carbon monoxide molecules, while \ce{CO2} has been found
to dominate \ce{CO} with a wide variety of mixing ratios in most comets
\citep[e.g.,][]{2010ApJ...717L..66O,2011Sci...332.1396A,2012ApJ...752...15O}.
Since CO is more volatile than \ce{CO2}, CO may be depleted in the
surface layers of comets after multiple passages through the inner solar
system, and the CO enrichment measured in \chris{} may indicate that of
the primordial material of the early solar nebula
\citep{2012ApJ...752...15O}.  We find that the evolution of the water
production rate is consistent with an estimate of the water vapor
outgassing from the subsolar point.  However, more sophisticated thermal
models considering the chemical differentiation of the nucleus and
emission from icy grains that were ejected from a localized region may
lead to slightly different profiles and smaller active area (Kossacki
and Szutowicz in preparation).

The thermal emission from the dust in the coma is detected by PACS
photometric observations in the blue and red channels. The asymmetry in
the red image indicates an anisotropic emission from the subsolar point
on the nucleus, consistent with the \ce{H2O} line blueshift.  Dust
production rates were computed for low (\SI{4e27}{\mols}) and high
(\SI{8e27}{\mols}) CO production rates and for the case of amorphous
carbon and olivine grains using the dust model described in
\citetalias{2010A&A...518L.149B}.  By comparing the flux density at the
optocenter in the blue and red bands with the dust thermal emission in
the model, we derived the dust production rates.  For a dust size
distribution index that explains the fluxes at the photocenters of the
\SIlist{70;160}{\micro\m} PACS images, the dust production rate is in
the range of \SIrange{70}{110}{\kg\per\s} at the time of the PACS
observations in August 2010, for the low and high outgassing cases.
These values of the production rate correspond to a dust-to-gas
production rate ratio of 0.3--0.4.

The dust production rates derived in the August 2010 observations are
roughly one order of magnitude lower than those in September 2009 from
the PACS observations \citepalias[see Table~\ref{tbl:qdust}
and][]{2010A&A...518L.149B}, while the CO production rate similarly
varies in the same time span assuming a post-perihelion trend $\propto
\rh^{-2}$, as measured in comet \object{C/1995 O1 (Hale-Bopp)}
\citep{1997EM&P...78....5B}.  This indicates that the dust-to-gas
production rate ratio remained approximately constant during the time
when the gas activity of the comet became increasingly dominated by CO
outgassing.  We find that the PACS data of August 2010 is best fitted by
a size index between \numlist{-4;-3}, consistent with the value derived
from the August 2009 PACS data \citepalias{2010A&A...518L.149B}.

\section{Conclusions}
\label{sec:conclusions}

Combining observations of rotational emission lines and dust thermal
emission provide valuable constraints on the properties of the gaseous
and dust activity in comets.  The \herschel{} Space Observatory was a
unique facility for obtaining sensitive observations of water emission
in comets at large heliocentric distances
\citep[e.g.,][]{2010DPS....42.0304B,2013A&A...560A.101O}, and imaging
the thermal dust coma in two channels simultaneously at far infrared
wavelengths \citep[see e.g.,
\citetalias{2010A&A...518L.149B};][]{2011ApJ...734L...1M}.

We presented HIFI and PACS observations of comet \chris{} that will be
an additional constraint in understanding the physical processes
responsible for the distant activity in comets. These observations will
complement previous measurements of \chris{} obtained with other
facilities at smaller heliocentric distances \citep[see e.g.,
\citetalias{2010A&A...518L.149B};][]
{2012ApJ...752...15O,2012Icar..220..277M,2013Icar..226..777R,2013DPS....4541318B}
that confirm that the activity of this object is dominated by
sublimation of molecular species more volatile than \ce{H2O}.  These
data will also add another set of data points to the available
observations of the distant activity of other comets such as
29P/Schwassmann-Wachmann~1 \citep{2010DPS....42.0304B} and C/1995 O1
(Hale-Bopp) \citep[e.g.][]{2002EM&P...90....5B,2003A&A...397.1109R}.

\begin{acknowledgements}
HIFI has been designed and built by a consortium of institutes and
university departments from across Europe, Canada, and the United States
under the leadership of SRON, Netherlands Institute for Space Research,
Groningen, The Netherlands, and with major contributions from Germany,
France, and the US.  Consortium members are: Canada: CSA, U.Waterloo;
France: CESR, LAB, LERMA, IRAM; Germany: KOSMA, MPIfR, MPS; Ireland, NUI
Maynooth; Italy: ASI, IFSI-INAF, Osservatorio Astrofisico di
Arcetri-INAF; Netherlands: SRON, TUD; Poland: CAMK, CBK; Spain:
Observatorio Astron\'omico Nacional (IGN), Centro de Astrobiolog\'ia
(CSIC-INTA). Sweden: Chalmers University of Technology -- MC2, RSS \&
GARD; Onsala Space Observatory; Swedish National Space Board, Stockholm
University -- Stockholm Observatory; Switzerland: ETH Zurich, FHNW; USA:
Caltech, JPL, NHSC.
PACS has been developed by a consortium of
institutes led by MPE (Germany) and including UVIE (Austria); KU Leuven,
CSL, IMEC (Belgium); CEA, LAM (France); MPIA (Germany);
INAF-IFSI/OAA/OAP/OAT, LENS, SISSA (Italy); IAC (Spain). This
development has been supported by the funding agencies BMVIT (Austria),
ESA-PRODEX (Belgium), CEA/CNES (France), DLR (Germany), ASI/INAF
(Italy), and CICYT/MCYT (Spain).
Support for this work was provided by NASA through an award issued by
JPL/Caltech.  We thank M.~S\'anchez-Portal from the \herschel{}
Science Center, and H.~Linz from the PACS Instrument Control Center
(ICC) for their help with the reduction and analysis of the \herschel{}
data.  The anonymous referee is thanked for providing constructive
comments and help in improving the contents of the paper.  M.dV.B.\
acknowledges partial support from grants NSF AST-1108686 and NASA
NNX12AH91H.  E.J.\ is FNRS Research Associate, D.H.\ is Senior Research
Associate and J.M.\ is Research Director FNRS.  C.O.\ thanks the Belgian
FNRS for funding her PhD thesis.  S.S.\ acknowledges support from Polish
MNiSW under grant 181/N-HSO/2008/0.  L.R.\ was supported by the Special
Priority Program 1488 (PlanetMag, \url{http://www.planetmag.de}) of the
German Science Foundation.
\end{acknowledgements}

\bibliographystyle{aa}
\bibliography{ads,preprints}

\end{document}